\documentclass{article}
\usepackage{frascatiphys,here,graphicx,subfigure}
\begin{document}
\title{ Scalar mesons from heavy flavor decays }
\author{Alberto Reis \\
{\em Centro Brasileiro de Pesquisas F\'{\i}sicas} \\
}
\maketitle
\baselineskip=11.6pt
\begin{abstract}
In the past few years the B-factories became unexpected players in the
scalar mesons business: in order to access the CP violation effects, it is
necessary to handle the dynamics of the strong interaction between the final state
hadrons. A number of large statistics studies heavy flavor decays involving a
scalar component have been performed recently by Belle and BaBar, who have joined
CLEOc, BES, E791 and FOCUS in the effort to understand the physics of the
scalar mesons. In this talk, the most recent results from these experiments  
will be reviewed, with emphasis on the low energy $K\pi$ system and on the f0(1370).
\end{abstract}
\baselineskip=14pt
\section{Introduction}
The identification of the scalar mesons is a long standing problem. There are too
many candidates  with mass bellow 2 GeV/c$^2$, although some states, like the
$\kappa$ and the $f_0(1370)$, are still controversial. In addition to the regular 
$q\bar q$ mesons, the observed spectrum may contain other types of states, 
like glueballs, hybrids, tetraquarks or molecules.

There are difficulties from both experimental and theoretical points of view. 
Scalars are, in general, broad overlapping states. Since there is no spin, they 
decay isotropically. In scattering experiments the production rates for low mass states,
like the $\sigma$ and $\kappa$, are suppressed by the Adler zeroes. At higher masses,
disentangling the broad, spinless states, like the $f_0(1370)$, from the smooth background is
complicated by the interference with other scalars and with the continuum,
which distorts the line shapes. 

The precise determination of pole positions and couplings to specific modes of all existing
scalar particles is an essential step towards the identification of the genuine $q\bar q$ states.
Such an ambitious task could not be accomplished by one single type of experiment. 
One has to look at the scalars problem from all possible perspectives, exploring
the different constraints imposed by different reactions.

In the past six or seven years there has been lots of new results on scalar mesons from
heavy flavor (HF) decays to light quarks (LQ), exploring the unique features of these
processes. In this paper we will focus on two crucial problems: the nature of the 
$\kappa$ meson and the existence of the $f_0(1370)$.  More
specifically, we will present the latest results from $\tau$ lepton and 
three-body decays of $D$ and $B$ mesons to light hadrons.

The paper is organized as follows. In Section 1 we discuss why HF decays are
a very useful tool for the study of scalar mesons. We will also discuss some aspects of
the basic analysis techniques in HF decays. In Section 2 we discuss the situation of 
the low mass
$K\pi$ spectrum. In Section 3 we discuss the issue of the $f_0(1370)$, in the light
of hadronic $D$ and $B$ decays. The last Section contains a summary and conclusions.

\section{Heavy flavor to light quarks}

There are unique features that make decays of $D$ and $B$ mesons to light hadrons 
very suitable for the study of scalar mesons. These particles 
are copiously produced in $D$ and $B$ decays, especially when there is a pair 
of identical particles in the final state. The B-factories already have very large charm 
samples, with a high degree of purity. Soon there will be 
also large samples of $B \to h_1h_2h_3$ decays ($h_i=\pi,K$) from the LHC experiments. 
With these high quality data, the $\pi\pi$ and $K\pi$ 
spectra can be accessed continuously, starting from threshold, and covering the entire 
elastic range.

Another appealing feature, especially in $D$ decays, is the close connection between 
the quark content of the initial state and the observed resonances. In the
decay of a $D$  meson the weak decay of the $c$  quark is embedded in a
strongly interacting system that leads to the final state hadrons. However, if one goes
through the PDG listing, one realizes that, in spite of the complexity of the $D$ decay, 
nearly the entire hadronic and semileptonic rates can be described by a rather simple 
scheme, in terms of 
tree-level valence quark diagrams, and the regular $q \bar q$ mesons from the
Constituent Quark Model. The dominant amplitudes in $D$ decays are the external
(spectator) and internal $W$ radiation. The 'final state valence quarks' result from
transition $c \to s(d) + u \bar d (\bar s)$ plus the 'spectator' $\bar q$. These quarks
define the possible intermediate $q \bar q$ states. This simple picture works very
well for intermediate states having either a vector, an axial-vector or a tensor 
resonance. If one excludes the scalar mesons, in $D$ decays
there is nothing else than the members of the usual $q \bar q$ 
nonets of the Constituent Quark Model. In other words, $D$ decays can be seen
as a '$q \bar q$ filter': if a resonance is observed in $D$ decays, then it is very 
likely to be a $q \bar q$ meson. One can expect that this holds also for the scalar
resonances, so $D$ decays would also provide clues about the nature of these mesons.

Semileptonic decays of the type $D \to h_1h_2l\nu$, and hadronic decays
of the $\tau$ lepton, $\tau \to h_1h_2\nu$, provide a 
particularly clean environment for the study of the scalar mesons, since
there is no strong interaction between the $h_1h_2$ pair and the leptons. However,
the $h_1h_2$ system is dominantly in P-wave in both cases. 
The S-wave contribution is typically less than 10\%, so very large samples
are required. An additional difficulty is the fact that the neutrinos are not 
reconstructed, so the background level in these decays is relatively large.

Hadronic decays of $D$ mesons, on the other hand, are much easier to be reconstructed. 
In some final states the S-wave component is largely dominant, like in the 
$D^+ \to K^- \pi^+\pi^+$ and in the $D^+,D^+_s \to \pi^-\pi^+\pi^+$ decays.
Background levels are typically of the order of a few percent. The problem here is how
to disentangle the desired information. The final states are strongly interacting
three-body systems, with a complex and unknown production dynamics. The pure $h_1h_2$ 
is certainly the main ingredient, but there is no direct route to extract it in
a model independent way. Approximations, and, therefore,
interpretation of the results, are unavoidable, unfortunately.

Most of the existing data on HF $\to$ LQ come from hadronic three-body $D$ decays. 
The event distribution in the Dalitz plot is given by,

\begin{equation}
\frac{d\Gamma}{ds_a ds_b} = \frac{\mathrm {1}}{32(2\pi M)^3} 
\mid \mathcal{M}(s_a,s_b) \mid ^2 ,
\end{equation}
where $M$ is the mass of the decaying particle and $s_a$, $s_b$ are the 
two-body invariant masses squared. The phase space density is constant,
so any structure in the Dalitz plot reflects
directly the dynamics of the decay.

The analysis technique of such decays is by now standard\cite{asner}.
The decay amplitude is written as a coherent sum of phenomenological amplitudes
corresponding to the possible intermediate states,
 
\begin{equation}
\mathcal{M}(s_a,s_b) = \left| \sum_L  ~f_D^L  ~\mathcal{S}^L 
~\mathcal{A}^L\right|^2,
\end{equation}
with

\begin{equation}
\mathcal{A}^L = \sum c_k e^{i\delta_k} A_k^L,   A_k^L  = f_R^L \times BW_k
\end{equation}

In the above equations $f_D^L$ and $f_R^L$ are form factors, with $L$ being the 
orbital angular momentum at the $D$ or at the resonance decay vertex; 
$\mathcal{S}^L$ is a function accounting for the angular momentum conservation, 
and $BW_k$ is a relativistic Breit-Wigner function (with an energy
dependent width, in general). The complex coefficients $c_k e^{i\delta_k}$ are usually the
fit parameters.

Analyzes differ by the way the S-wave, $\mathcal{A}^0$, is modeled. There are
three basic approaches.

The most used model is the so called isobar model\cite{e791dp,cg}, in which the S-wave
is represented by a sum, of Breit-Wigner functions and a nonresonant amplitude, 
with unknown complex coefficients. The nonresonant amplitude is assumed to be uniform 
in $D$ decays, but varies across the Dalitz plot in the case of $B$ decays. 
This model is simple and intuitive, but 
there are well known conceptual problems with this approach\cite{meissner}.
As we will see, it provides an effective description of the data but, in some cases, the
physical interpretation of the results is rather ambiguous.

Another approach is the K-matrix model\cite{sm}. This is a very sophisticate tool, but
it is based on a very strong assumption: in the $D$ decay, the resonant $h_1h_2$ system 
and the bachelor particle recoil against each other without any interaction. 
This would greatly simplify the problem, although one must acknowledge the lack of 
experimental evidence supporting this assumption. If the three-body rescattering is negligible, 
the evolution of the $h_1h_2$ system must be the same as in $h_1h_2$ elastic scattering. 
In the K-matrix approach the S-wave amplitude is, therefore, fixed by other type of reactions,
whereas the production of the $h_1h_2$ pair is an unknown function to be determined by the fit.
This universality of the S-wave is often seen as the most appealing feature of the K-matrix 
approach. We should not forget that the same constraint should be applied to all partial waves,
which, unfortunately, is not the case of the  analyzes published so far. 
In general, good fits are achieved only if an extra energy-dependent phase 
is added. In the framework of the K-matrix model, the origin of such a phase is attributed 
to the production of the $h_1h_2$ pair. This is simply a matter of interpretation, since
this phase could also be due to the rescattering of the final state particles. There is no
way, with the existing measurements, to distinguish between these two effects.

Finally, there is the PWA approach\cite{bm}. Here no assumption is made
about the content of the S-wave, which is treated as a generic
complex function of the $h_1h_2$ mass, $\mathcal{A}^0 = a(s)e^{i\phi (s)}$. 
The $h_1h_2$ spectrum is divided in bins. At each bin edge the amplitude is defined by
two real constants, $a_k$ and $\phi_k$, which are fit parameters. The amplitude at  
any value of the $h_1h_2$ mass is given by a polynomial interpolation. This method 
relies on a precise modeling of the P- and D-waves. The problem resides, once more, in the
interpretation of the results. The measured S-wave phase $\phi (s)$ includes 
any rescattering/production effects, which should be disentangled in order to
determine the 'bare' $h_1h_2$ amplitude.

One last remark is in order. A consequence of representing the decay amplitude
by a coherent sum of amplitudes is that the decay fractions do not add to 100\%.
Different amplitudes populate the same region of the phase space, so they are expected to interfere. 
The amplitudes have phases that vary across the Dalitz plot. The interference can be
destructive in some regions of the phase space and constructive in other ones. One should 
be very careful, though, when the sum of fractions largely exceeds 100\%. In almost all
cases this large interference occur between the broad amplitudes in the S-wave, involving, 
in general, the nonresonant amplitude. This is a symptom of a poor modeling of the
S-wave. In other words, one may find a mathematical solution to the fit problem, but
with a misleading physical interpretation. In the following
Sections we will see several instances of this problem.

\section{The low energy $K\pi$ spectrum: the $\kappa$ problem}

There are two crucial questions to be answered in the low mass $K\pi$ spectrum.
The neutral $\kappa$ has been observed by different experiments
in several types of heavy flavor decays\cite{cg,kmat,beskappa}, 
but so far evidence for its charged partner has been
scarce and controversial. This casts doubts about the $\kappa$ being a 
regular $I=1/2$ $q\bar q$ state. 
The other issue is the pole position of the $\kappa$. There is no data on $K\pi \to K\pi$
elastic scattering bellow 825 GeV/c$^2$. Determination of the $\kappa$ pole 
position\cite{descortes} must rely on extrapolations of the existing data. New data 
that fill the existing gap would be highly desirable. This section is devoted to these
two issues.

\subsection{The $D^0 \to K^+K^-\pi^0$  decay  --  BaBar}

\begin{figure}[H]
\begin{center}
 {\includegraphics[scale=0.38]{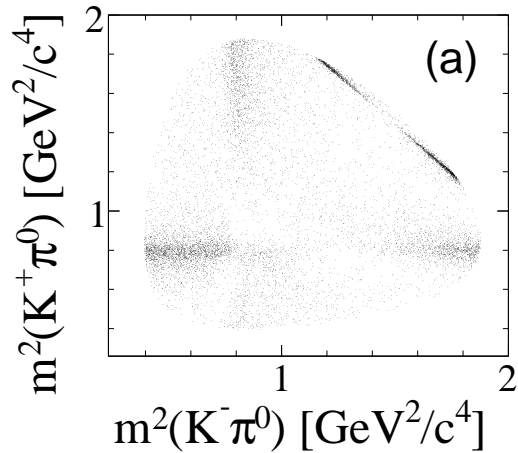}}
\caption{\it Dalitz plot of the $D^0 \to \overline{K}^0 \pi^+ \pi^+$ decay\cite{babar1}. The narrow
band at the upper edge of the Dalitz plot correspond to the $\phi\pi^0$ channel.}
\label{kpipi01}
\end{center}
\end{figure}

In a recent analysis of the decay $D^0 \to K^+K^-\pi^0$, performed by the BaBar collaboration\cite{babar1}, 
the issue of the charged $\kappa$ has been addressed. The data sample has 11K signal
events with 98\% purity. The Dalitz plot of this decay is shown in Fig. \ref{kpipi01}.
Resonances can be formed in all three axes (the third axis, $K^-K^+$, is along the diagonal, 
starting at high $K^{\pm}\pi^0$ mass) of the Dalitz plot. 

\begin{figure}[H]
\begin{center}
 {\includegraphics[scale=0.29]{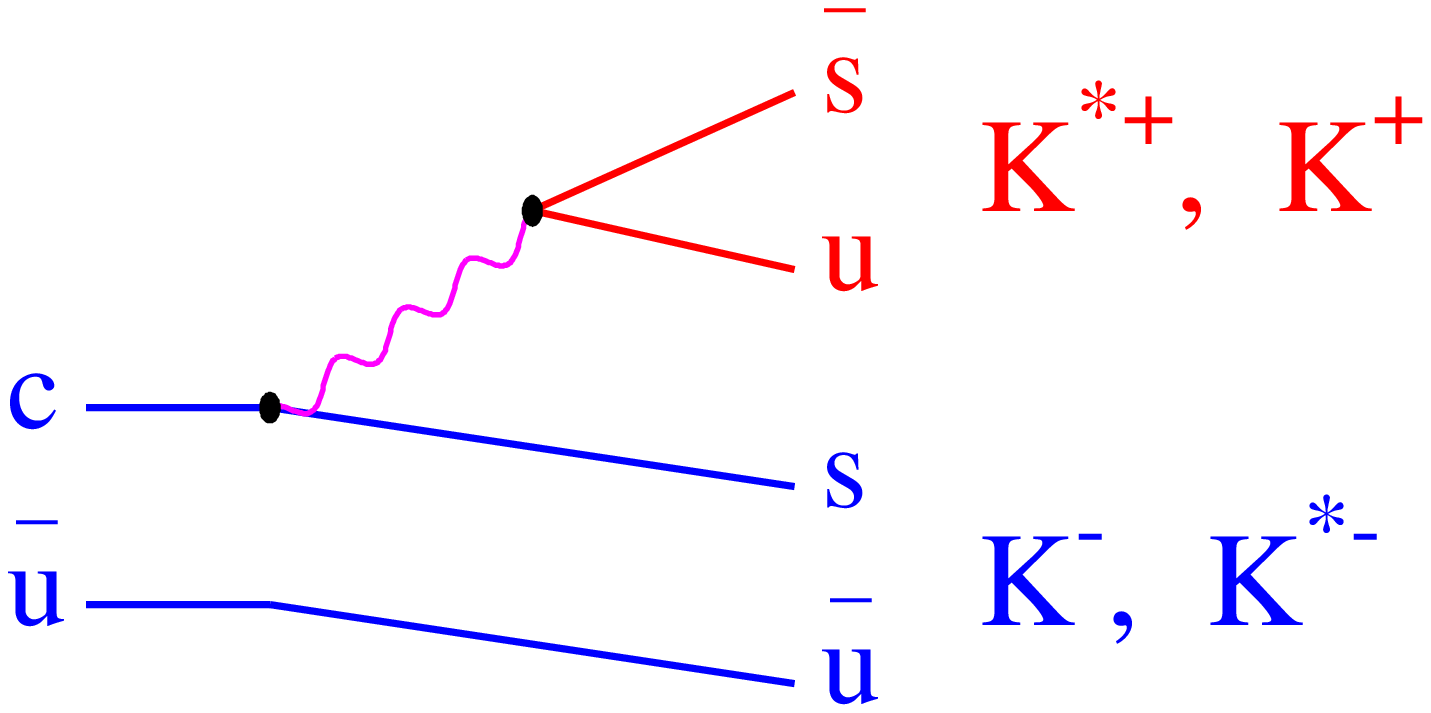}}
 {\includegraphics[scale=0.29]{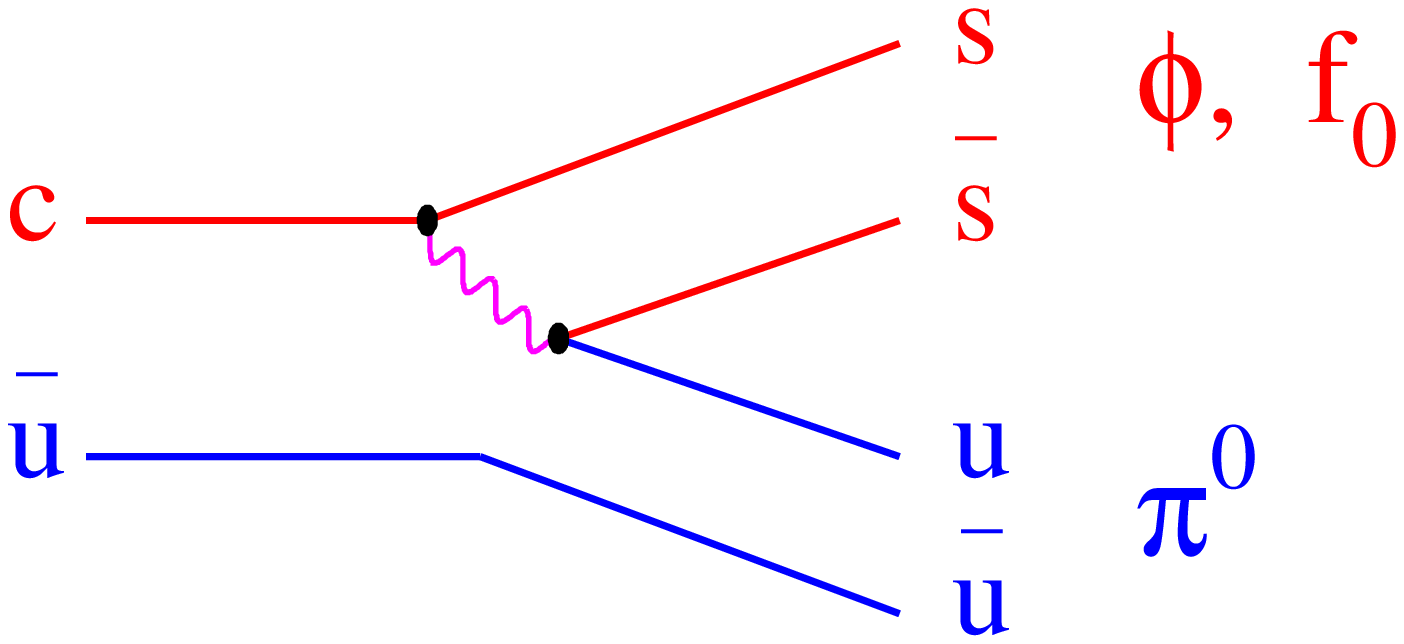}}
\caption{\it Valence quark diagrams leading to the $K^+K^-\pi^0$ final state: external (left) and
internal (right) $W$ radiation. The
resonances that can be formed by the 'final state' quarks are indicated.}
\label{diag2}
\end{center}
\end{figure}

The main diagrams for this decay are shown in Fig. \ref{diag2}.  
There is a well defined pattern in $D$ decays that proceed via the external $W$-radiation amplitude
(Fig. \ref{diag2}-a): due to the $V$-$A$ nature of the weak interactions, the $W$ couples much more
strongly to a vector or an axial-vector meson than to a pseudoscalar. According to this
pattern, one expects a dominant contribution from the $K^*(892)^+K^-$ channel, 
compared to $K^*(892)^-K^+$. Important contributions from the $K^*(892)^-K^+$ and $\phi \pi^0$ modes
are also expected. We can see in Fig. \ref{kpipi01} clear structures corresponding 
to these resonances.

The Dalitz plot of Fig. \ref{kpipi01} was fitted using three different models for the $K^{\pm}\pi^0$ S-wave. 
In the first fit, the S-wave was represented by the usual isobar model -- a constant nonresonant amplitude 
plus two Breit-Wigner functions for the $\kappa^{\pm}\pi^0$ and $K^*_0(1430)^{\pm}\pi^0$ modes --
with parameters taken from E791\cite{cg}. The second fit used the E791 PWA S-wave\cite{bm} measured from
the $D^+ \to K^-\pi^+\pi^+$ decay. Finally, the third
fit used the LASS $I=1/2$ $K^-\pi^+$ S-wave amplitude\cite{lass}. The isobar model has the smaller fit
probability ($\chi^2$ prob. $<$ 5\%). The E791 PWA S-wave provides a good description of the data
($\chi^2$ prob. $=$ 23\%), but the best fit was obtained using the LASS $I=1/2$ S-wave. Results of the
fits with LASS $I=1/2$ S-wave are summarized in Table \ref{kkpi0}. Note that in Model II
the exclusion of the $K^*(1410)K$ amplitudes 
-- a 5\% contribution in Model I -- has a minor impact on the other P-wave components,
but causes a dramatic change in the S-wave fraction.
Moreover, in Model II the sum of the decay fractions largely exceeds 100\%, indicating 
the existence of large destructive interference effects.

\begin{table}[t]
\centering
\caption{ \it Decay fractions (\%) of the $D^0 \to K^+K^-\pi^0$ decay. Results are from
fits using the $I=1/2$ LASS S-wave amplitude.}
\begin{tabular}{|l|c|c|} \hline
    mode      &  model I & model II \\
\hline
\hline
$K^*(892)^+K^-$   &  45.2$\pm$0.9  &  44.4$\pm$0.9    \\
$K^*(1410)^+K^-$  &  3.7$\pm$1.5   &	    -	      \\ 
$K^+\pi^0 (S) $   &  16.3$\pm$0.1  &  71.1$\pm$4.2    \\
$\phi \pi^0$      &  19.3$\pm$0.7  &  19.4$\pm$0.7    \\
$f_0(980)\pi^0$   &  6.7$\pm$1.8   &  10.5$\pm$1.4    \\ 
$K^*(892)^-K^+$   &  16.0$\pm$0.9  &  15.9$\pm$0.9    \\  
$K^*(1410)^-K^+$  &  2.7$\pm$1.5   &	    -	      \\     
\hline
$\chi^2$ prob.    &     62\%       &     47\%         \\
\hline
\end{tabular}
\label{kkpi0}
\end{table}

The fit with the LASS amplitude is much better than the one having explicitly the $\kappa^{\pm}$ 
amplitude.Note that this result does not rule out the charged $\kappa$, since its pole can be found 
in LASS data\cite{descortes}. It simply says that the isobar representation of the $K^-\pi^+$ S-wave 
in the $D^+ \to K^-\pi^+\pi^+$ decay, from E791\cite{cg} and FOCUS\cite{kmat} analysis, 
is not a good model for the $K^+\pi^0$ S-wave amplitude in the  $D^0 \to K^+K^-\pi^0$ decay.
In both E791 and FOCUS analyzes of the $D^+ \to K^-\pi^+\pi^+$ decay, 
the LASS $I=1/2$ S-wave amplitude fails to provide a good description of the data, whereas
a very good fit was achieved with the isobar model. The origin of this discrepancy is yet to be 
understood. With a larger $D^0 \to K^+K^-\pi^0$ sample, a model independent measurement of the 
S-wave amplitude could be performed and directly compared to the results of the E791 MIPWA\cite{bm}. 

\subsection{The $\tau^- \to \overline{K}^0\pi^- \nu_{\tau}$ decay --  Belle}

In hadronic decays of the $\tau$ lepton, $\tau \to h_1h_2\nu_{\tau}$, the $h_1h_2$ system is not affected 
by strong interaction with the leptonic current. This would be as close as one could get to the
$h_1h_2 \to h_1h_2$ elastic scattering using HF decays. The Belle Collaboration published 
recently\cite{belletau} a study of the decay $\tau^- \to \overline{K}^0\pi^- \nu_{\tau}$. 
A sample with 53K signal events was selected from the reaction $e^+e^- \to \tau^+\tau^-$.
The $\tau^+$ decays to a muon plus two neutrinos, so the signature of the event is one lepton recoiling
against three charged prongs in the opposite hemisphere. With a total of three
missing neutrinos, it is very difficult to reconstruct the event topology. Immediate consequences
are a high background level ($\sim$20\% in this analysis) and the lack of an angular analysis.

The angular distribution of the helicity angle was used in FOCUS study\cite{massa} of the 
$D^+ \to K^-\pi^+ \mu^+ \nu$ decay. The helicity angle is formed by the $K^-$ momentum and 
the line of flight of the $D^+$, measured in the $K^-\pi^+$ rest frame. In this decay there is a 
5\% scalar component. The line shape of the $K^-\pi^+$ spectrum, which is dominated by the 
$K^*(892)^0$, is not sensitive to the different models for the S-wave component. 
However, the different possibilities -- a complex constant, a Breit-Wigner function, or the 
LASS  $I=1/2$ S-wave amplitude -- lead to different angular distributions, so one can explore this
feature in order to understand the nature of the scalar component.

\begin{figure}[H]
\begin{center}
 {\includegraphics[scale=0.29]{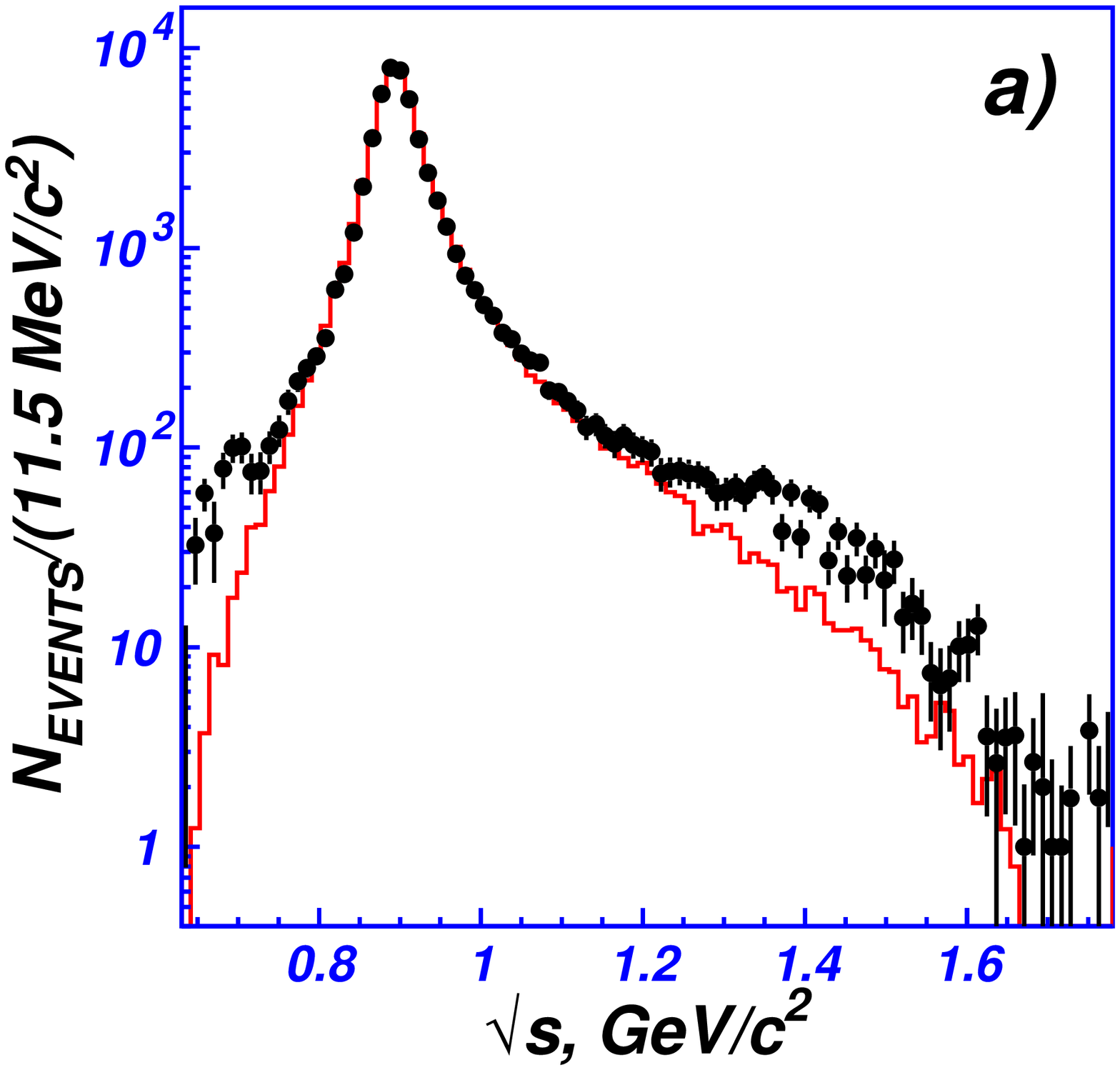}}
 {\includegraphics[scale=0.29]{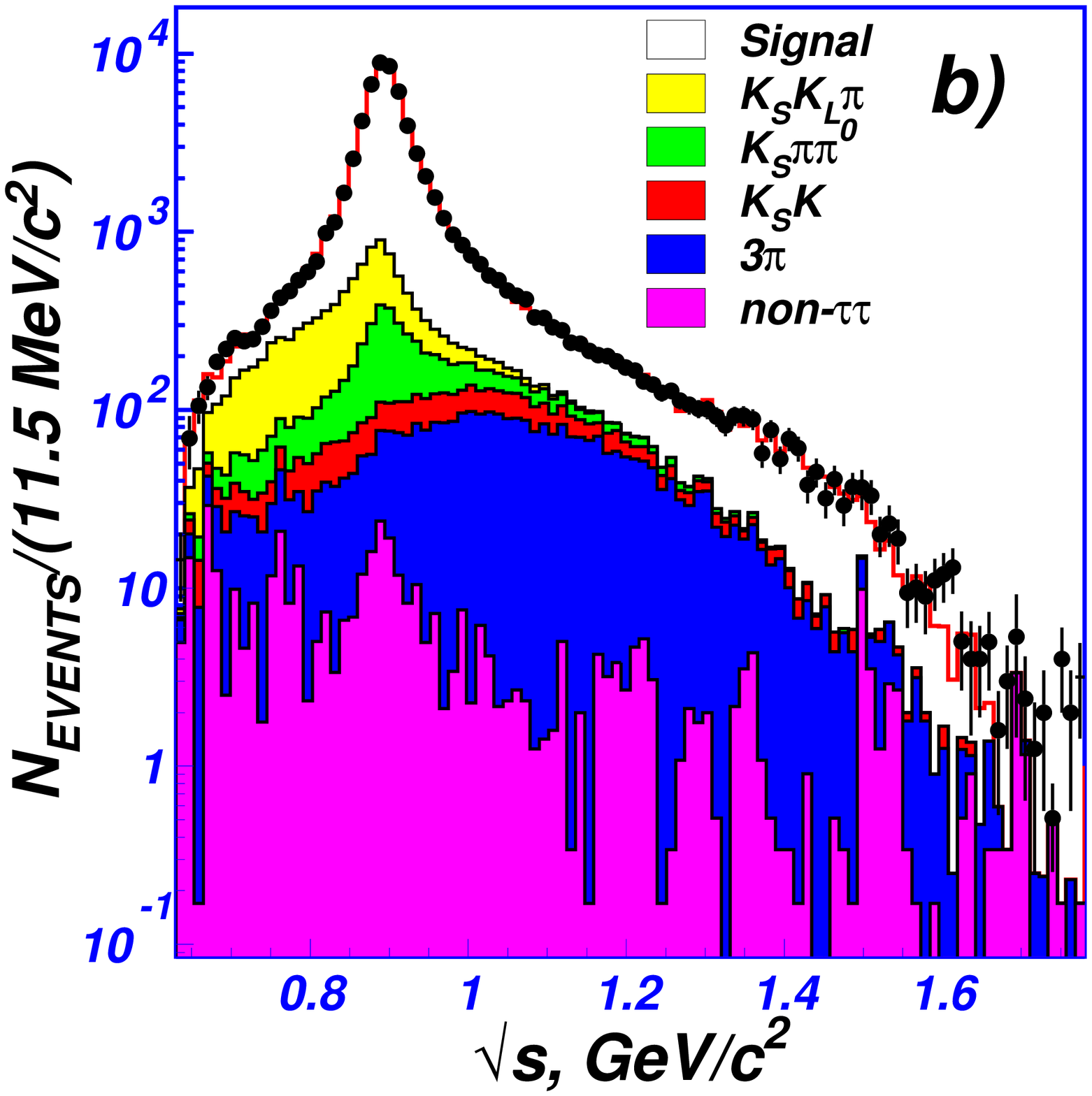}}
\caption{\it a) The  $\overline{K}^0\pi^-$ mass spectrum\cite{belletau} (points with error bars). The histogram
represents the $K^*(892)^-$ contribution. b) The $\overline{K}^0\pi^-$ mass spectrum with
the different background components and the result of the best fit superimposed.}
\label{belletau}
\end{center}
\end{figure}

In the case of the $\tau^- \to \overline{K}^0\pi^- \nu_{\tau}$ decay, the $K^*(892)^-$ 
alone is not enough
to describe the $\overline{K}^0\pi^-$ spectrum, as we can see in Fig. \ref{belletau}-a. 
There is an excess of events in the lower and in the upper part of the spectrum. The 
$\overline{K}^0\pi^-$ spectrum was fit with two classes of models : the dominant  $K^*(892)^-$
plus the charged $\kappa$ and one high mass state (either a vector or a scalar resonance);
the  $K^*(892)^-$ plus the LASS $I=1/2$ $K^-\pi^+$ S-wave amplitude. 
The best fit was obtained with the $K^*(892)^-$  plus a pure S-wave, that is,
a sum of the $\kappa$  and the $\overline{K}^*_0(1430)^-$. The confidence level of this fit is 41\%
(Fig. \ref{belletau}-b). The fit using the LASS amplitude, on the other hand, yielded a C.L. 
of only 10$^{-8}$.

The result of Belle analysis is in conflict with those of the BaBar study of the $D^0 \to K^+K^-\pi^0$ 
decay. One could expect that, in the absence of strong interactions, the 
$\overline{K}^0\pi^-$ S-wave phase would match that of LASS, whereas in the case of $D^0 \to K^+K^-\pi^0$
the rescattering would cause deviations from the elastic phase. This is a rather intriguing result.
In both cases the statistics is limited to a few thousands of events, though. The angular analysis  
on a larger $\tau^- \to \overline{K}^0\pi^- \nu_{\tau}$ sample may confirm the
resonant behavior of the S-wave at low mass. In this case we would have
a compelling evidence of the charged $\kappa(800)$.

\subsection{The $D^+ \to K^-\pi^+ \pi^+$  decay --  FOCUS}
The $D^+ \to K^-\pi^+ \pi^+$ is a golden mode for the study of scalar mesons. The S-wave
accounts for over 60\% of the decay rate. In addition, 
the large branching fraction (9.5\%), combined with the long $D^+$ lifetime  and a final state having
only charged tracks (the odd charge track being always a kaon), makes the $D^+ \to K^-\pi^+\pi^+$ 
the easiest charm decay mode to be reconstructed. Using this decay one can
measure the $K^-\pi^+$ S-wave amplitude near threshold, filling the existing gap in LASS data.

FOCUS has published recently a study of this decay using the K-matrix approach\cite{kmat}. A good fit was 
obtained combining the $I=1/2$ and $I=3/2$ LASS S-wave phases with  an additional energy 
dependent phase. This extra phase was interpreted as being originated from the production dynamics of the 
$K\pi$ system, but it could also be attributed to the rescattering of the final state particles.
One important aspect should be stressed. The $I=3/2$ component is purely nonresonant, whereas all resonances
are in the $I=1/2$ amplitude. The latter has also a nonresonant background. The fractions of each isospin 
component in the K-matrix fit are (207$\pm$28)\% and (40$\pm$ 10)\% for the $I=1/2$ and $I=3/2$, respectively. 
The total S-wave fraction amounts to (83$\pm$2)\%. Here is another instance of large interference between
broad components within the S-wave.  

The PWA approach for the  $K^-\pi^+$ S-wave was used recently by CLEOc\cite{cleoc} and FOCUS\cite{eu}.
In this new FOCUS study, the $D^+ \to K^-\pi^+ \pi^+$ Dalitz plot is fitted using also the isobar model 
for the $K^-\pi^+$ S-wave. This is a work in progress, based on a sample of 93K signal events and with 
98\% purity. From here to the end of this Section we will discuss the preliminary results this FOCUS 
PWA/isobar Dalitz plot analysis.

In the isobar fit, the S-wave model was that of E791\cite{cg}: a sum of an uniform
nonresonant amplitude, plus two Breit-Wigner functions for the $\kappa \pi^+$ and  
$\overline{K}^*_0(1430) \pi^+$ modes. In order to make a direct comparison with E791 and CLEOc, 
the S-wave Breit-Wigner functions are multiplied by the same Gaussian form factors, $f_D=e^{-p^{*2}r_D^2/12}$.
Masses and widths of the scalar resonances are fit parameters.

The fit fractions and resonance parameters
are shown in Table \ref{isob}. Results from the three experiments are in reasonable
agreement, except for the S-wave fraction in CLEOc analysis, which is a bit smaller than E791 and FOCUS.
But the most remarkable feature in Table \ref{isob} is shown in the last three columns. The isobar fit to FOCUS
data was repeated with a little variation of the S-wave parameterization. In FOCUS(b) the value of the scalar 
form factor parameter, $r_D$, was set to 6 GeV$^{-1}$ (instead of 5 GeV$^{-1}$ in FOCUS(a)). In FOCUS(c) 
the value of $r_D$ was set to zero, which is equivalent of having no scalar form factor.  All the three FOCUS 
fits have an equally good confidence level. We observe a dramatic change in the S-wave composition, while the 
P- and D- waves remain unaltered.  The nonresonant fraction varies by a factor of four.

\begin{table}[t]
\centering
\caption{ \it Decay fractions (\%) of the $D^+ \to K^-\pi^+\pi^+$ decay. Results are from
fits using the isobar model for the S-wave amplitude. The masses and widths of scalar resonances are in
units of MeV/c$^2$.}
\begin{tabular}{|l|c|c|c|c|c|} \hline
 mode                          & E791          &   CLEOc       &  FOCUS(a)     & FOCUS(b)       & FOCUS(c) \\
\hline
\hline
$\overline{K}^*(892)^0\pi^+$   & 12.3$\pm$1.4  & 11.2$\pm$1.4  &  11.3$\pm$0.3 &  11.7$\pm$0.3   & 11.2$\pm$0.3  \\ 

$\overline{K}^*(1410)^0\pi^+$  &  -            & -             &  1.2$\pm$0.3  &  1.1$\pm$0.3	 & 1.3$\pm$0.3   \\ 

$\overline{K}^*(1680)^0\pi^+$  & 2.5$\pm$0.8   & 1.4$\pm$0.2   & 3.3$\pm$0.3   &  2.7$\pm$0.3	 & 3.8$\pm$0.3   \\ 

$\overline{K}^*_2(1430) \pi^+$ &  0.5$\pm$0.2  & 0.4$\pm$0.4   & 0.20$\pm$0.05 &  0.20$\pm$0.05  & 0.20$\pm$0.05 \\ 
                         
$\overline{K}^*_0(1430) \pi^+$ & 12.5$\pm$1.4  & 10.5$\pm$1.3  & 16.8$\pm$0.8  &  14.3$\pm$0.7   & 18.7$\pm$1.2  \\ 

$\kappa(800) \pi^+$            & 47.8$\pm$13.2 & 31.2$\pm$3.6  & 43.5$\pm$4.5  &  71.3$\pm$5.5   & 22.3$\pm$3.2  \\ 

$\mathrm{nonresonant}$         & 10.4$\pm$1.4  & 13.0$\pm$7.3  & 14.3$\pm$3.0  &  7.5$\pm$3.1	 & 31.6$\pm$4.5  \\ 
\hline 
$\kappa(800)$ mass             & 797$\pm$48    & 805$\pm$11    & 837$\pm$12    &  829$\pm$14    & 867$\pm$14    \\
$\kappa(800)$ width            &  410$\pm$97   & 453$\pm$21    & 443$\pm$21    &  433$\pm$18    & 485$\pm$27    \\
$\overline{K}^*_0(1430)$ mass  & 1461$\pm$3    & 1459$\pm$5    & 1466$\pm$4    &  1468$\pm$4    & 1466$\pm$4    \\
$\overline{K}^*_0(1430)$ width & 169$\pm$5     & 175$\pm$16    &  193$\pm$7    &  193$\pm$7     &  192$\pm$7    \\
\hline
\end{tabular}
\label{isob}
\end{table}

\begin{figure}[H]
\begin{center}
 {\includegraphics[scale=0.38]{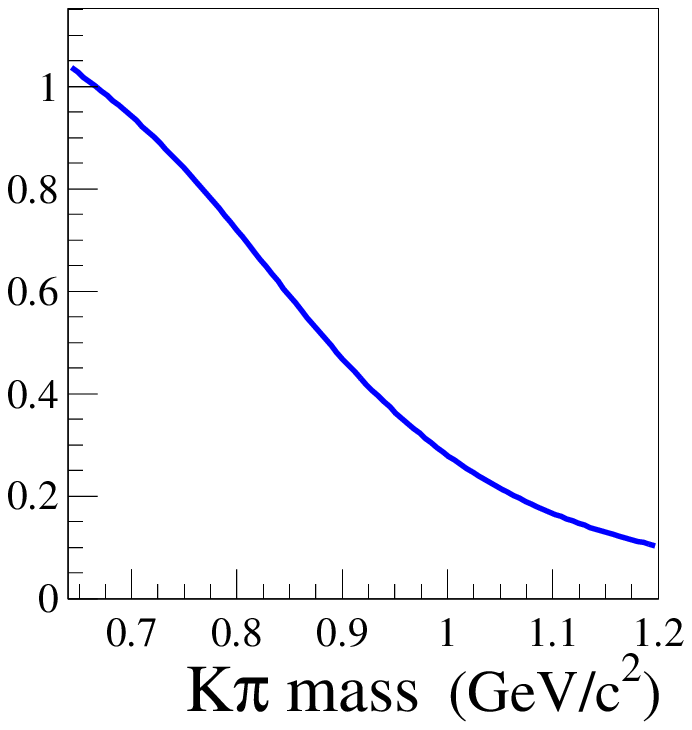}}
 {\includegraphics[scale=0.38]{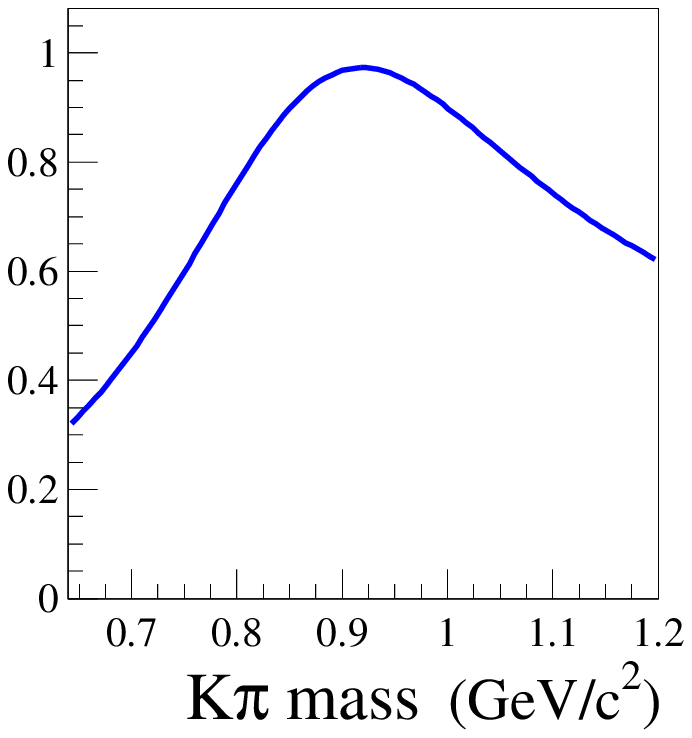}} 
 {\includegraphics[scale=0.38]{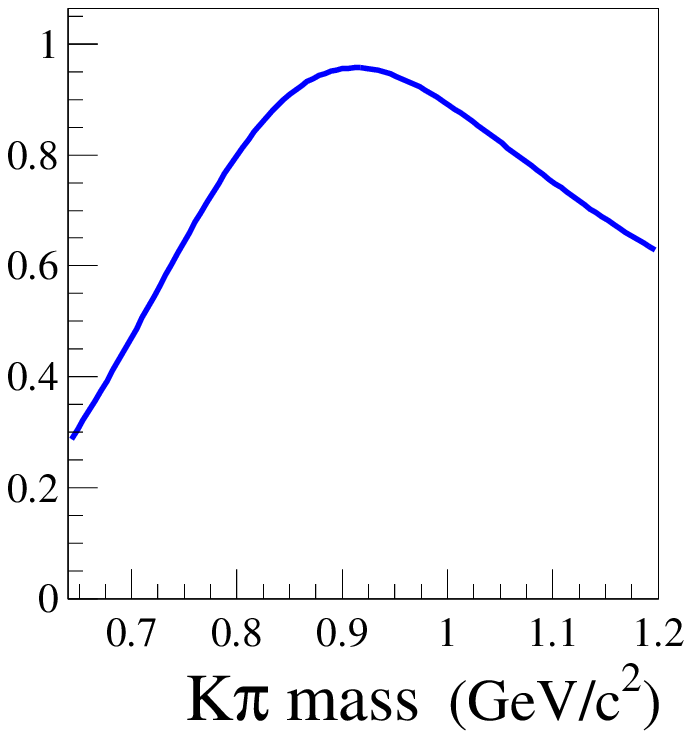}}
\caption{\it Left: The modulus squared of the $\kappa$ Breit-Wigner, without the scalar form factor; center:
the modulus squared of the $\kappa$ Breit-Wigner multiplied by the scalar form factor; right: the modulus 
squared of the sum of the $\kappa$ Breit-Wigner, without the scalar form factor, and a complex constant.}
\label{kappabw}
\end{center}
\end{figure}

This instability can be readily explained by  the interference between the broad $\kappa$ Breit-Wigner and the uniform
nonresonant component. In Fig. \ref{kappabw} the effect of the scalar form factor is illustrated. Fig. \ref{kappabw}-a
shows the modulus squared of the $\kappa$ Breit-Wigner. The $K\pi$ mass dependence of the width, in the denominator of 
the Breit-Wigner, shifts the maximum of the function to a value bellow the $K\pi$ threshold. The introduction of the 
Gaussian form factor modifies the line shape near threshold (Fig. \ref{kappabw}-b). The resulting  $\kappa$ amplitude 
has a more reasonable behavior. The same effect can be achieved if, instead of multiplying the $\kappa$ 
Breit-Wigner by the Gaussian form factor, we add the correct amount of a constant nonresonant amplitude, with the 
right phase difference. This is shown in Fig. \ref{kappabw}-c. 

The case of the $D^+ \to K^-\pi^+\pi^+$ is didactic: the isobar model yields a good fit, but it fails to
provide an unambiguous physical picture of the decay dynamics. The model for the S-wave is clearly inadequate.
We need to go beyond the isobar model in order to understand the composition of the S-wave.

The $D^+ \to K^-\pi^+\pi^+$ Dalitz plot is also fitted using the PWA method. The $K\pi$ spectrum is divided in 
forty equally spaced intervals. At the edge of each interval the S-wave amplitude is defined by two fit parameters,
$\mathcal{A}_0(s=s_k)=a_ke^{i\phi_k}$. The value of the S-wave amplitude at any point in the  $K\pi$ spectrum is given
by a spline interpolation of these forty points. The S-wave is determined by an iterative procedure. A first fit
is performed fixing the P-wave to that of the isobar analysis. The S-wave is determined (80 fit parameters). 
A second fit is performed fixing now the S-wave and varying the P-wave. The process is repeated until it converges.

\begin{figure}[H]
\begin{center}
{\includegraphics[scale=0.40]{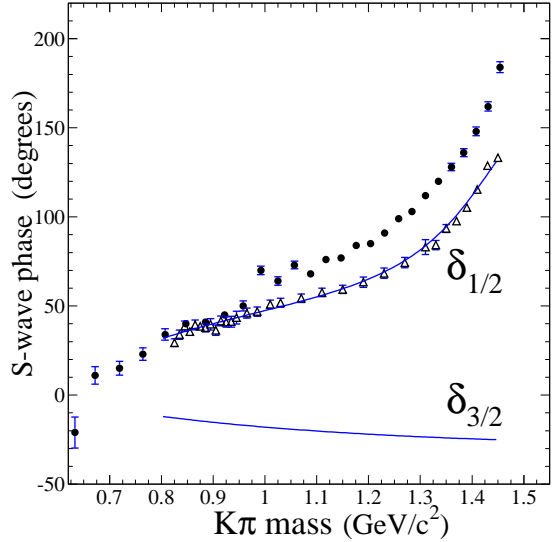}}
\caption{\it The phase of the S-wave $K\pi$ amplitude, from the $D^+ \to K^-\pi^+\pi^+$ decay, as a function
of the $K\pi$ mass squared. The phase was shifted up by 80 degrees, in order to make a comparison with
the LASS $I=1/2$ S-wave phase, $\delta_{1/2}$. The LASS $I=3/2$ S-wave phase, $\delta_{3/2}$, is also shown.}
\label{kpppwa}
\end{center}
\end{figure}

Fig. \ref{kpppwa} shows the FOCUS PWA S-wave phase, $\phi(s)$, as a function of the $K\pi$ mass squared. 
A 80$^o$ overall phase was added to $\phi(s)$ for a better comparison with the $I=1/2$ and $I=3/2$ phases from
LASS. There is a clear discrepancy between the S-wave phase measured by FOCUS and that of LASS $I=1/2$ amplitude. 
The agreement between FOCUS and LASS cannot be achieved even combining both isospin phases. An extra phase, which 
depends smoothly on the $K\pi$ mass, is necessary for the matching of the two amplitudes. The origin of such a phase 
is unclear.

The PWA is as close as one can get to a model independent method, since no hypothesis is assumed for the S-wave.
But the method has some technical limitations arising from the large number of free parameters. One limitation is the 
existence of multiple solutions. With so many parameters, the fit has freedom to cure eventual problems with the 
P-wave model. A precise representation of the P-wave is mandatory, otherwise some 'leakage' into the S-wave would be 
unavoidable.

The interpretation of the results is not trivial. The amplitude measured with the PWA method includes not only the 
$K\pi$ dynamics, but also any possible contribution from the production of the $K\pi$ pair and rescattering of the 
final state particles. Some urgent input from theory is required in order to disentangle these effects.

\section{Is there an $f_0(1370)$?}

In the low energy $\pi\pi$ spectrum the $\sigma$ pole is now well established, but
the situation between 1-2 GeV/c$^2$ is still controversial. There may be three neighbor
scalar states, namely the $f_0(1370)$, the $f_0(1500)$ and  the $f_0(1710)$. The
$f_0(1500)$ is a narrow, well established state, with mass, width and couplings know
to a good degree of accuracy\cite{pdg}, but the uncertainty on the $f_0(1370)$ parameters is
very large: $1200 < m_0 < 1500$ GeV/c$^2$ and $200 < \Gamma_0 < 500$ GeV/c$^2$. The existence of
this state is often questioned. 

The $f_0(1370)$ has been observed mostly in central production and $p\bar p$
annihilation (see, for instance D. Bugg's recent review on this state 
\cite{bugg}). BES has also reported on this state from the decay $J/\psi \to \phi \pi\pi$
\cite{besf0}. This state is very difficult to be detected because it is broad and very close 
to the $f_0(1500)$. Its line shape may also be sensitive to the opening of the $\sigma \to 4pi$ channel.

An interesting and related issue: the mass of the lightest scalar glueball is expected to be 
around 1.5 Gev/c$^2$. It is widely accepted that the three scalars would mixed states,
having both $q\bar q$ and $gg$ components in their wave functions.

Heavy flavor decays are particularly useful here, since in these decays the $q\bar q$
component of the intermediate resonances are probed.

\subsection{The $D^+_s \to \pi^- \pi^+ \pi^+$  decay  --  FOCUS and E791}

\begin{figure}[H]
\begin{center}
 {\includegraphics[scale=0.32]{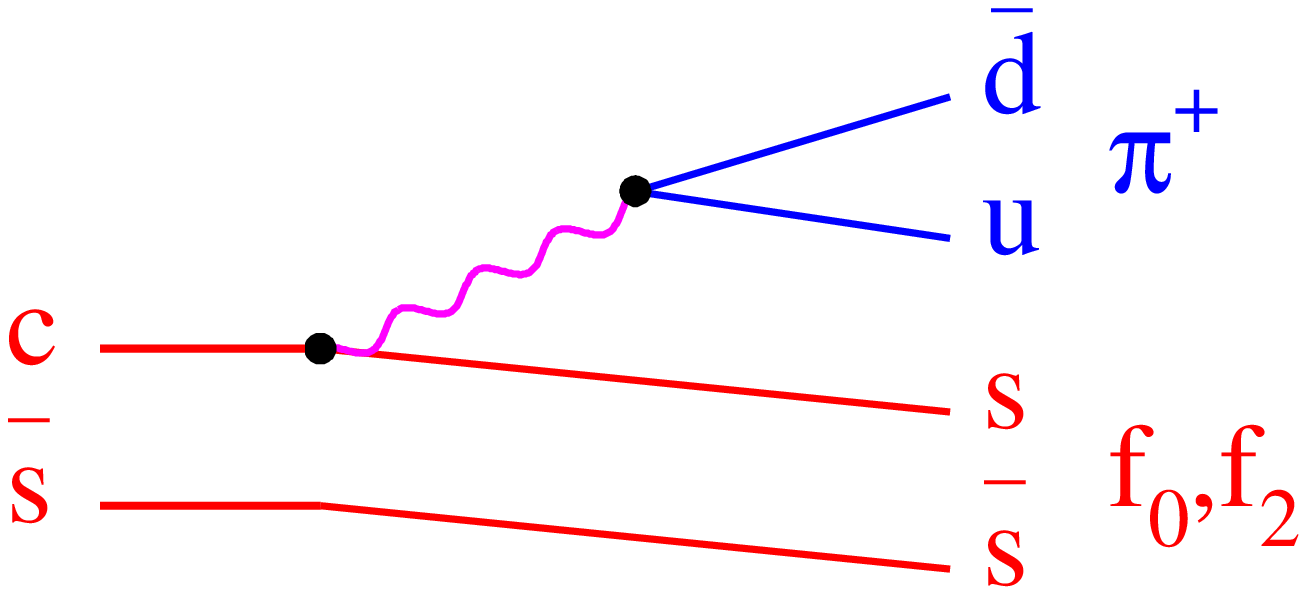}} 
 {\includegraphics[scale=0.25]{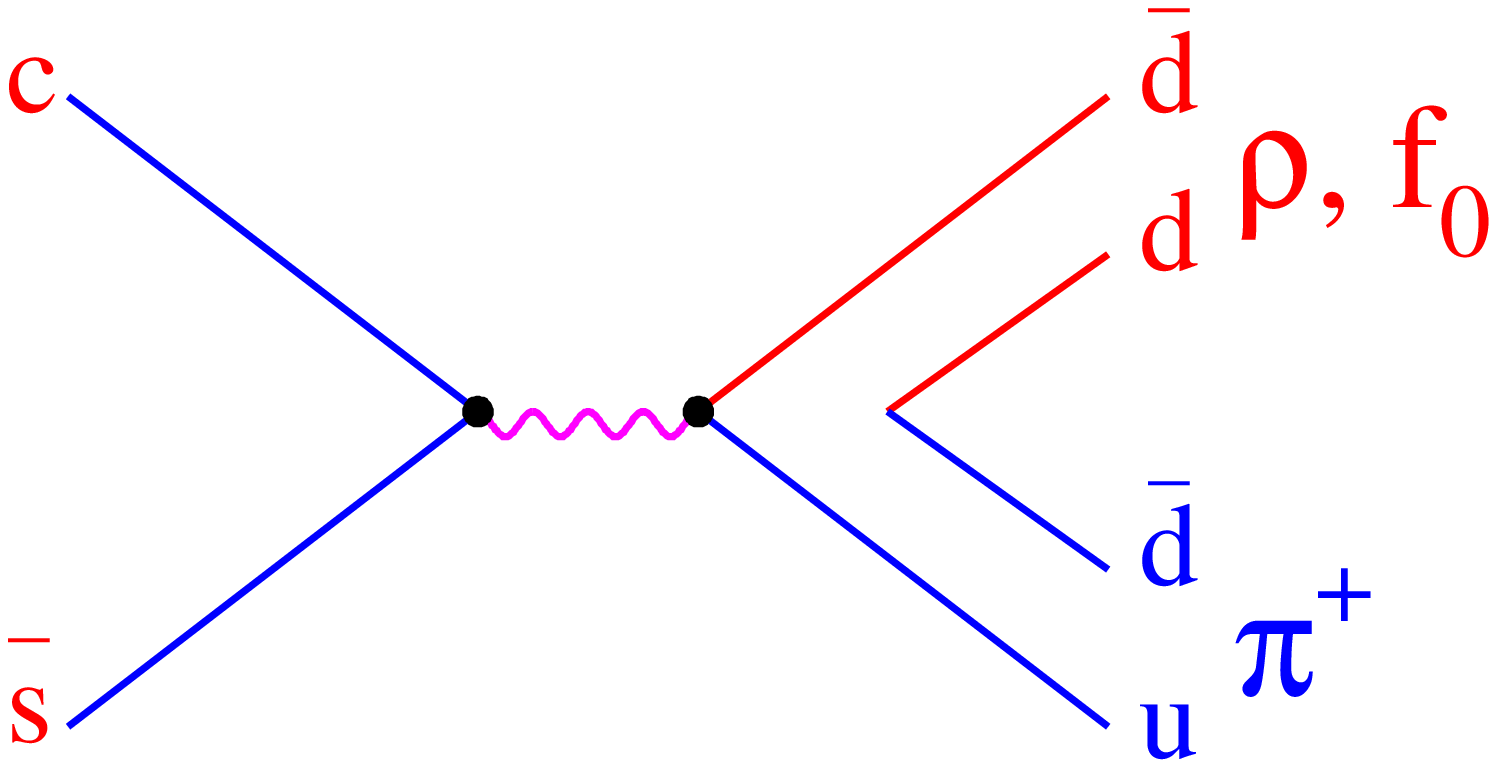}}
\caption{\it The dominant diagrams leading to the $D^+_s \to \pi^- \pi^+ \pi^+$ decay.}
\label{3pidiag}
\end{center}
\end{figure}

The $D^+_s \to \pi^- \pi^+ \pi^+$ decay is particularly suited to the study of scalar mesons. 
In this decay the S-wave component amounts to over 80\% of the 
decay rate. The dominant diagram is shown in Fig. \ref{3pidiag}-a, with a small contribution from the
annihilation diagram (Fig. \ref{3pidiag}-b). The $f_0(980)\pi^+$ should be the dominant mode. The $f_0(980)$ has a 
large coupling to $K^+K^-$ and, therefore, it must have a strong $s \bar s$ component in its wave function. 
An important contribution from the channel $f_2(1270)\pi^+$ is also expected.

\begin{figure}[H]
\begin{center}
 {\includegraphics[scale=0.38]{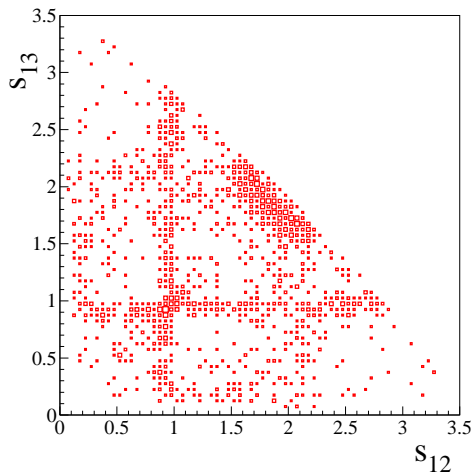}}
\caption{\it The Dalitz plot of the $D^+_s \to \pi^- \pi^+ \pi^+$ decay, from FOCUS.}
\label{3pidp}
\end{center}
\end{figure}

FOCUS has collected a sample of 1400 $D^+_s \to \pi^- \pi^+ \pi^+$ signal events. The Dalitz plot,
shown in Fig. \ref{3pidp}, was fitted with the K-matrix approach\cite{sm}. It was also fitted using
the isobar model, but the results of this analysis were not published.

There are two remarkable features in Fig. \ref{3pidp}:  a very clear structure at 1 (GeV/c$^2$)$^2$, 
corresponding to the $f_0(980)$; the concentration of events near the border, at 
$m_{\pi\pi}^2 \sim$ 2 (GeV/c$^2$)$^2$, which is due to the $f_2(1270)$, $\rho(1450)^0$ and to a scalar
state, which we refer to as $f_0(X)$. 
The Dalitz plot was fit\cite{dsdp} using the isobar model for the S-wave, which has three components: the
$f_0(980)\pi^+$, the $f_0(X)\pi^+$ and the nonresonant modes. The mass and width of the $f_0(X)$ 
are fit parameters.

The fit fractions, as well as the mass and width of the $f_0(X)$, are shown in Table \ref{dsisob} 
(FOCUS errors are statistical only). FOCUS has about twice as many events as E791. In both analysis
the $f_0(980)\pi^+$ is the dominant component, followed by the $f_0(X)\pi^+$. The BES Collaboration\cite{besf0}
measured the mass and width of the $f_0(1370)$  in the $J/\psi \to \phi \pi\pi$ decay, obtaining
$m_0 = (1350\pm50)$ GeV/c$^2$ and $\Gamma_0=(265\pm40)$ GeV/c$^2$. Comparing these values to the ones
of the $f_0(X)$, we conclude that the state observed in the $D^+_s \to \pi^- \pi^+ \pi^+$ is not the same as 
the one observed by BES.

Several variations of the S-wave model were tested.  The $\sigma \pi^+$ mode was added to the S-wave, but its
contribution is consistent with zero. A fit including the $f_0(1370)\pi^+$ (with BES parameters) was performed, 
yielding a null contribution of this mode.
In both FOCUS and E791, only one $f_0(X)$ is necessary to describe the data, and this state is consistent
with being the $f_0(1500)$. The decay fraction of the $f_0(X)$ is very large. Given the 
diagram of Fig. \ref{3pidiag}-a, one may conclude that this state has a significant $s \bar s$ component in 
its wave function. In this case we may also expect a large fration of this mode in the 
$D^+_s \to \pi^- \pi^+ \pi^+$ decay.

\begin{table}[t]
\centering
\caption{ \it Decay fractions (\%) of the $D^+_s \to \pi^-\pi^+\pi^+$ decay. Results are from
fits with the isobar model for the $\pi^-\pi^+$ S-wave amplitude.}
\begin{tabular}{|l|c|c|} \hline
 mode                              & FOCUS          &   E791\cite{e791ds}  \\
\hline
\hline
$f_0(980)\pi^+$                    & 76.9$\pm$4.9  &  56.5$\pm$5.9  \\
$f_0(X)\pi^+$                      & 23.3$\pm$0.5  &  32.4$\pm$7.9  \\
$\mathrm{nonresonant}$             & 13.2$\pm$5.7  &     1$\pm$2    \\
$\rho(770)^0\pi^+$                 &  1.2$\pm$0.1  &   5.8$\pm$4.4  \\
$\rho(1450)^0\pi^+$                &  4.0$\pm$1.0  &   4.4$\pm$2.1  \\
$f_2(1270)\pi^+$                   &  9.7$\pm$1.4  &  19.7$\pm$3.4  \\ 
\hline
$m_0(f_0(X))$ (GeV/c$^2$)          & 1.476$\pm$5.7 & 1.434$\pm$18   \\
$\Gamma_0(f_0(X))$ (GeV/c$^2$)     &  119$\pm$18   & 173$\pm$32     \\
\hline
\end{tabular}
\label{dsisob}
\end{table}

\subsection{The $D^+_s \to K^- K^+ \pi^+$ decay  --  BaBar}

\begin{figure}[H]
\begin{center}
 {\includegraphics[scale=0.25]{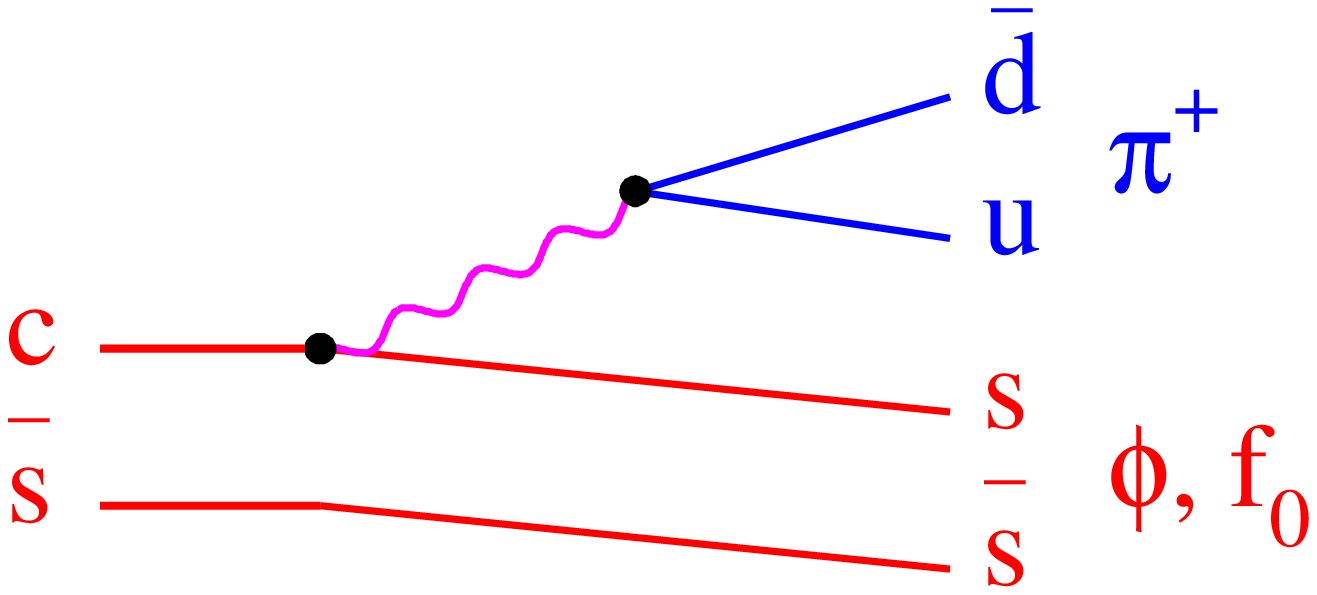}} 
 {\includegraphics[scale=0.25]{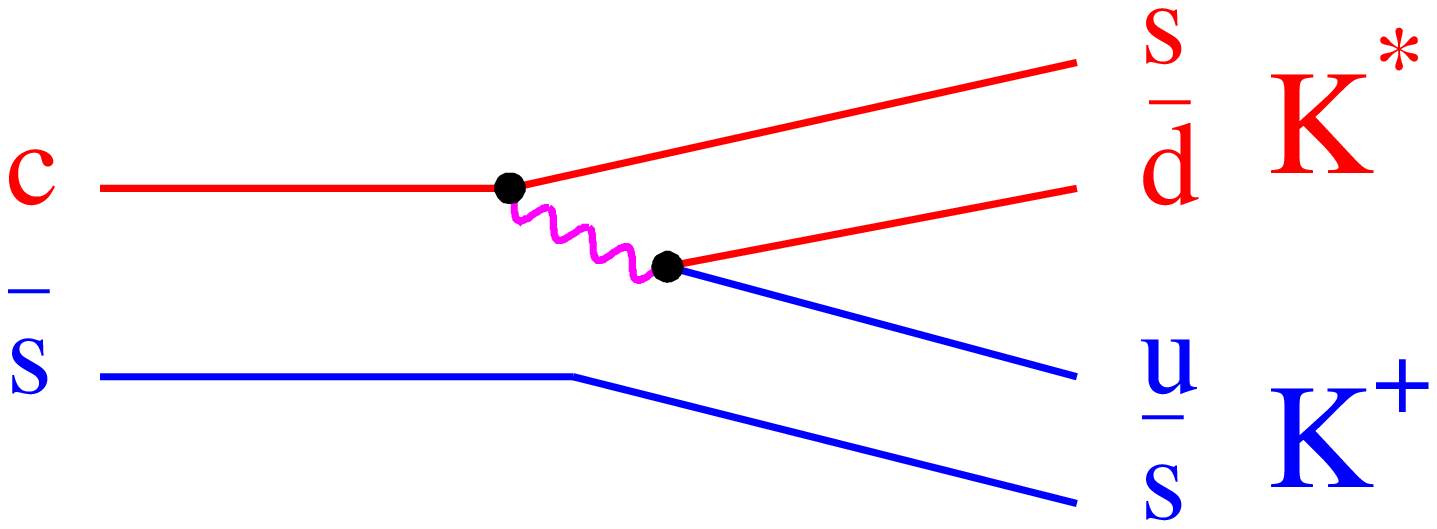}}
\caption{\it The dominant diagrams for the decay $D^+_s \to K^- K^+ \pi^+$.}
\label{kkpdiag}
\end{center}
\end{figure}

The  $D^+_s \to \pi^- \pi^+ \pi^+$ and the $D^+_s \to K^- K^+ \pi^+$ decays share the same
dominant diagram (Fig. \ref{3pidiag}-a). The latter mode can also proceed via the internal $W$-radiation
amplitude (Fig. \ref{kkpdiag}-b). A dominant contribution from the $\phi \pi^+$, 
$f_0(980) \pi^+$ and $f_0(X) \pi^+$ modes is expected, but there should also be a large 
$\overline{K}^*(892)^0 K^+$ component. 

\begin{figure}[H]
\begin{center}
 {\includegraphics[scale=0.45]{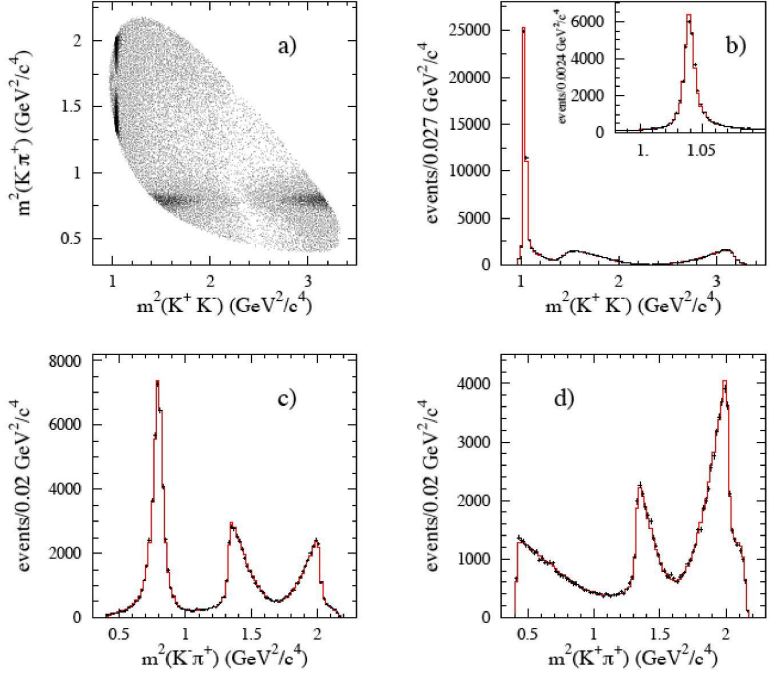}} 
\caption{\it a) The Dalitz plot of the $D^+_s \to K^-K^+\pi^+$ decay\cite{babarkkp}. Plots b) to d) show the
projections of the Dalitz plot into the three axes (points with error bars) with the fit result
superimposed (solid histograms).}
\label{kkpdp}
\end{center}
\end{figure}

BaBar collected a very large (100K signal events) and clean  (95\% purity)
sample\cite{babarkkp} of the decay $D^+_s \to K^- K^+ \pi^+$. The Dalitz plot is shown in Fig. \ref{kkpdp}.
BaBar is currently analyzing this data using the PWA method for the S-wave. Here we present preliminary results 
of the Dalitz plot analysis using the isobar model. 

The fit result is shown in Table \ref{dskkp}. In Fig. \ref{kkpdp} we have the Dalitz plot projections with the fit
result superimposed. In the low $K^+K^-$ mass region there could also
be a contribution from $a_0(980)\pi^+$, in addition to the $f_0(980)\pi^+$ and the $\phi\pi^+$. 
In practice, is nearly impossible to fit the data with a model having, at the same time, these three amplitudes. 
The interference between them is very large, the coefficients become highly correlated and the individual fractions
become too unstable. A stable fit is obtained with a model having only one of the two scalar amplitudes. The
values reported here are from a fit with  the $f_0(980)\pi^+$. Note that there is still a large uncertainty in 
the $f_0(980)\pi^+$ fraction.

\begin{table}[t]
\centering
\caption{ \it Decay fractions (\%) of the $D^+_s \to K^-K^+\pi^+$ decay, from a
fit using the isobar model.}
\begin{tabular}{|l|c|} \hline
 mode                              & fraction(\%)            \\
\hline
\hline
$f_0(980)\pi^+$             &   35$\pm$14    \\
$f_0(1370)K^+$              &  6.3$\pm$4.8   \\
$f_0(1710)K^+$              &  2.0$\pm$1.0   \\
$\phi(1020)\pi^+$           & 37.9$\pm$1.9   \\
$\overline{K}^*(892)^0K^+$  & 48.7$\pm$1.6   \\
\hline
\end{tabular}
\label{dskkp}
\end{table}

The most surprising result is the absence of the $f_0(X)\pi^+$. We can see in  Fig. \ref{kkpdp}-b
that there are very few events in the $f_0(X)$ region. There is, on the other
hand, a small excess of events next to the $\phi / f_0(980)$ region, which is not well described by an
uniform nonresonant amplitude. Instead, a scalar state was introduced. The fitted mass and width of this 
state are $m_0 = (1.313 \pm 10 \pm 114)$ GeV/c$^2$ and $\Gamma_0 = (0.395 \pm 8 \pm 133)$ GeV/c$^2$.
The large errors reflect the sensitivity to the details of the S-wave parameterization. One could not really 
interpret this result as an indication of the $f_0(1370)$, since this is a very complicated region of the
$K^+K^-$ spectrum.

The absence of a $f_0(X)$ contribution in $D^+_s \to K^-K^+\pi^+$ may indicate that the S-wave model used 
in the study of the $D^+_s \to \pi^-\pi^+\pi^+$  is not the most correct. BaBar is 
currently analyzing a sample of the decay $D^+_s \to \pi^-\pi^+\pi^+$ which is a factor of 10
larger than that of FOCUS. The most important result would be a simultaneous PWA measurement of the
S-wave in both final states. The statistics is not a problem for the $D^+_s \to K^-K^+\pi^+$,
but, unfortunately, it is still a bit limited in the case of $D^+_s \to \pi^-\pi^+\pi^+$, even considering 
the BaBar sample. 

\subsection{The $D^0 \to \overline{K}^0 \pi^+ \pi^+$ decay  --  Belle}

Belle and BaBar have collected very large samples of the $D^0 \to \overline{K}^0 \pi^+ \pi^-$ decay.
The Dalitz plot analysis performed by both experiments used the isobar model, and the results
are in very good agreement. Here we will discuss the Belle analysis, based on a sample of 
534K events\cite{bellekspipi} with 98\% purity.

\begin{figure}[H]
\begin{center}
 {\includegraphics[scale=0.25]{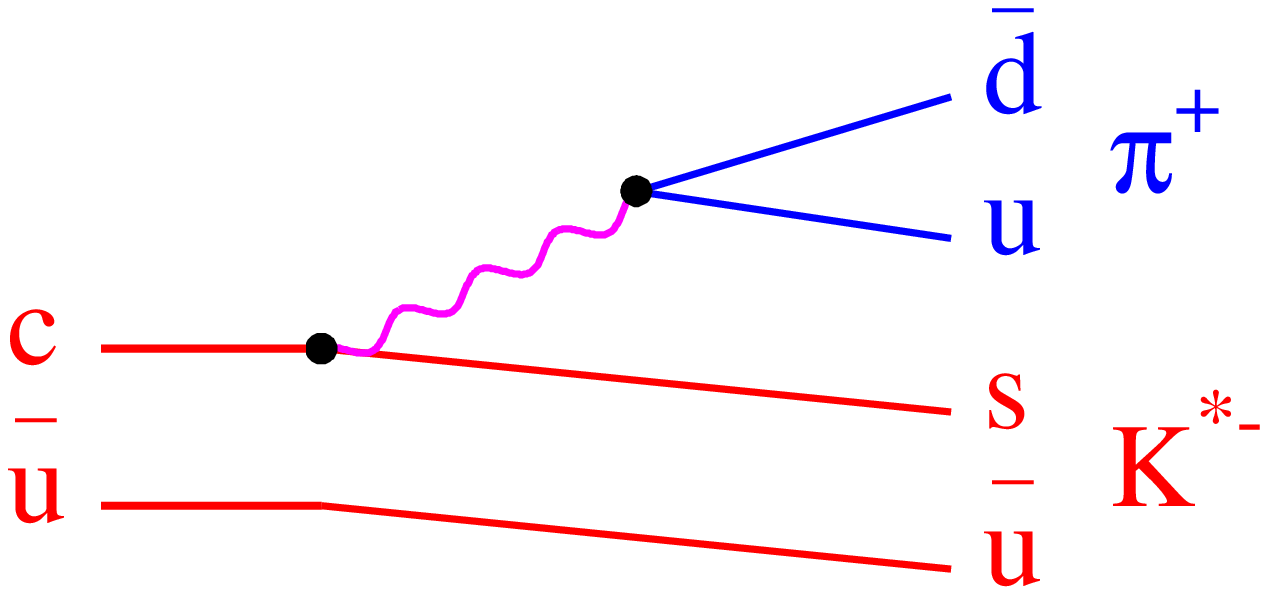}} 
 {\includegraphics[scale=0.25]{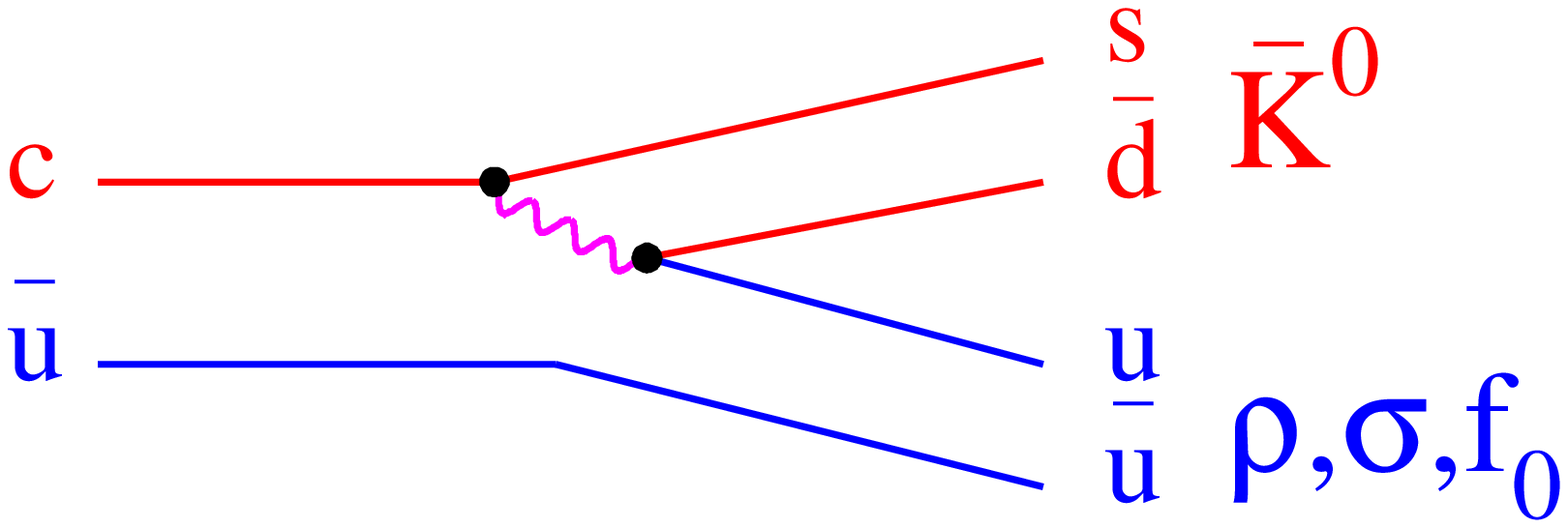}}
\caption{\it Diagrams for the $D^0 \to \overline{K}^0 \pi^+ \pi^-$ decay. }
\label{ksppdiag}
\end{center}
\end{figure}

\begin{figure}[H]
\begin{center}
 {\includegraphics[scale=0.35]{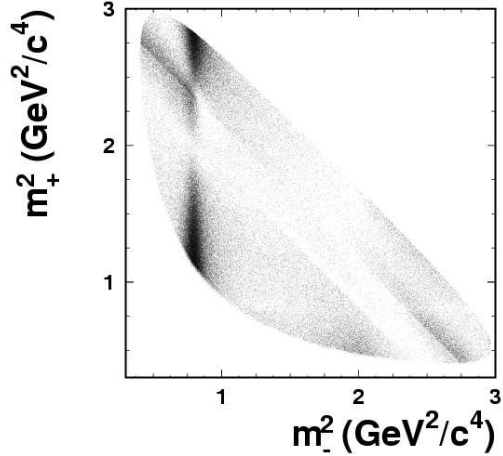}} 
\caption{\it The Dalitz plot of the decay $D^0 \to \overline{K}^0 \pi^+ \pi^-$\cite{bellekspipi}. 
The label $m_-^2$ refer to
the $\overline{K}^0 \pi^-$ mass squared, when the parent is a $D^0$, and to the $K^0 \pi^+$ mass squared,
when the parent is a $\overline{D}^0$.}
\label{ksppdp}
\end{center}
\end{figure}

The diagrams for this decay are shown in Fig. \ref{ksppdiag}. The dominant amplitude should be
the $K^*(892)^-\pi^+$ channel, with important contributions from the $\rho(770) \overline{K}^0$ mode
and form the $\pi^-\pi^+$ S-wave. The Dalitz plot, shown in Fig. \ref{ksppdp}, is very complex, since there
could be resonances in all three axis. Moreover, there is a small contribution from the
doubly Cabibbo suppressed decay $D^0 \to K^0 \pi^+ \pi^-$. In Fig. \ref{ksppdp} the label $m_-^2$ refer to
the $\overline{K}^0 \pi^-$ mass squared, if the parent is a $D^0$, or to the the $K^0 \pi^+$ combination, 
in case the parent is a $\overline{D}^0$.

The fit model has 19 amplitudes
leading to the $\overline{K}^0 \pi^+ \pi^-$ final state. The $\pi^+\pi^-$ S-wave contains four amplitudes: 
$\sigma \overline{K}^0$, the $f_0(980)\overline{K}^0$, $f_0(X)\overline{K}^0$ and an extra scalar
state, the $\sigma_2 \overline{K}^0$. The $f_0(X)$ parameters were taken from E791 (see Table \ref{dsisob}),
whereas the $\sigma_2$ parameters were determined by the fit. This extra $\sigma_2$ amplitude was introduced
to account for a structure near $m_{\pi\pi}\sim$1 GeV/c$^2$, but it does not correspond to a real state. The 
parameters obtained by the fit are $m_0= (1.059 \pm 6)$ GeV/c$^2$ and $\Gamma_0= 0.059 \pm 10$ GeV/c$^2$.

The dominant contribution is, as expected, the $K^*(892)^-\pi^+$ (62\%), followed by the
$\rho^0 \overline{K}^0$ (21\%) and the $\pi^+\pi^-$ S-wave ($\sim$15\%). The contribution of the
$f_0(X)$ is small but significant (1.6\%). No errors on the fractions were quoted.

In spite of the large number of amplitudes, including the extra '$\sigma_2$', a good fit was not obtained.
It is very hard to obtain a good C.L. in fits to very large samples. The goodness-of-fit is accessed by
$\chi^2$-like tests, in which the phase space is divided in bins of variable area, so that the number of
events is similar in all bins. The isobar approach is, perhaps, too simplistic. With such a large sample,
relatively small and localized deviations from the observed Dalitz plot distribution have, in general, large 
impact in the fit C.L.

The goal of this analysis is to study the mixing phenomenon. The technique is a time dependent Dalitz plot
analysis. For this purpose, an effective representation of the data would suffice, according to the authors. 
Unfortunately the treatment given to the $\pi^+\pi^-$ S-wave do not allow us to draw any conclusion.
We cannot interpret the fraction attributed to the $f_0(X)$ as an evidence for this state, since the $\pi\pi$ 
S-wave is not well understood. No attempt to measure its parameters was reported. A study of the
$D^0 \to \overline{K}^0 \pi^+ \pi^-$ decay focused at the $\pi\pi$ S-wave and using the PWA method is in order.

\subsection{Charmless hadronic three-body decays of $B$ mesons.}
Charmless hadronic three-body decays of $B$ mesons are a very promising tool, but there is still
a long way to go. The data samples resemble those of $D$ mesons from the late 80's. There are two main problems: 
statistics is still limited and the background is still high. The nonresonant component is another problem.
It is likely to be larger in $B$ than in $D$ decays. A constant nonresonant amplitude is the usual parameterization
in the case of $D$ decays, which may be a good approximation given the limited phase space.
In $B$ decays, however, the understanding of the nonresonant amplitude is a crucial issue\cite{bed}, as one 
can already conclude from the existing data.

There has been intense activity in this area, with many studies from the B-factories. Here we will focus 
on two analyzes from Belle, the $B^+ \to K^+ \pi^+ \pi^-$\cite{belle2} and $B^0 \to K^0 \pi^+ \pi^-$\cite{belle3} 
decays, and on two analysis by BaBar, the $B^+ \to K^+ K^+ K^-$\cite{babarkkk1} and 
$B^0 \to K^0 K^+ K^-$\cite{babarkkk2} decays.

\begin{figure}[H]
\begin{center}
 {\includegraphics[scale=0.29]{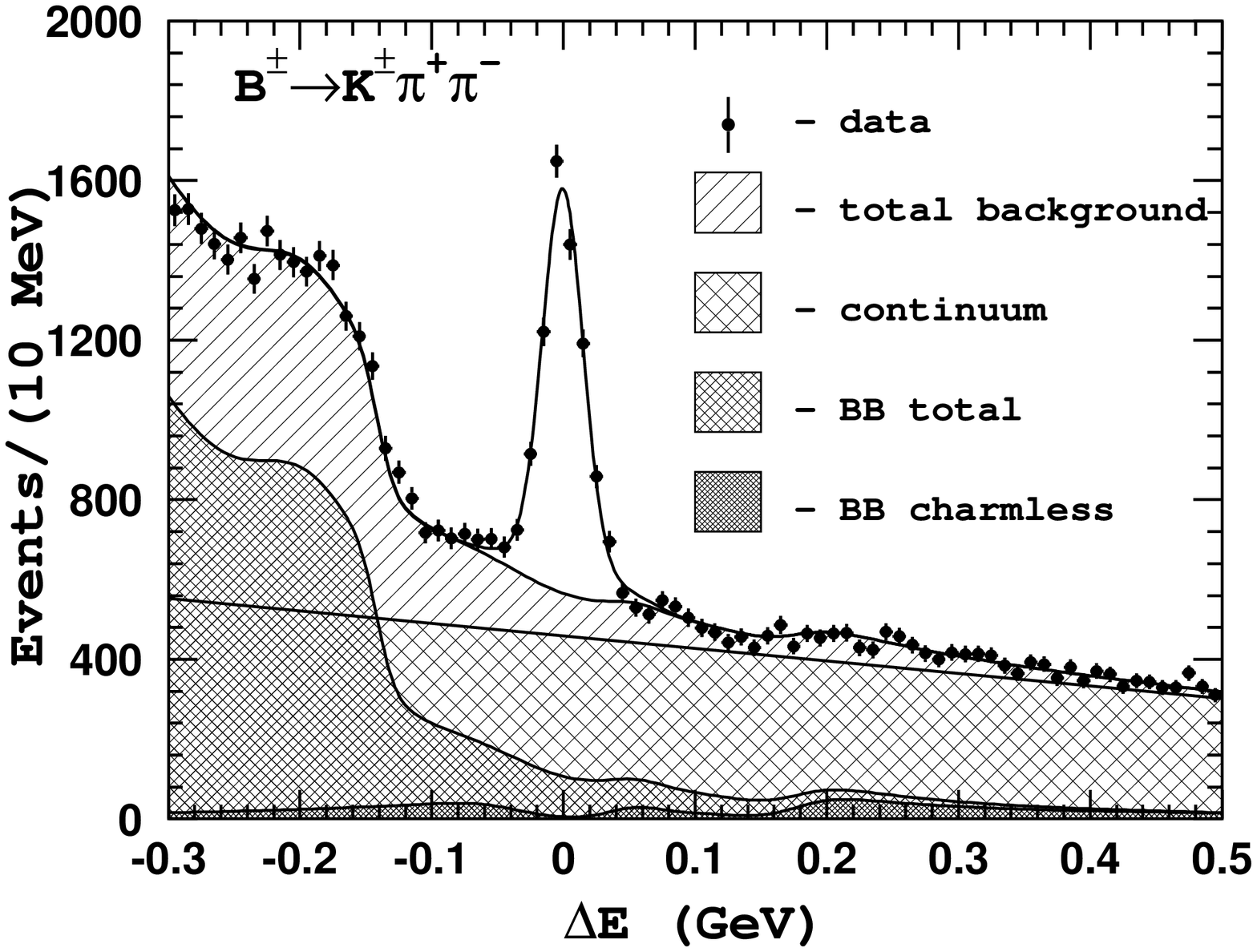}}
 {\includegraphics[scale=0.25]{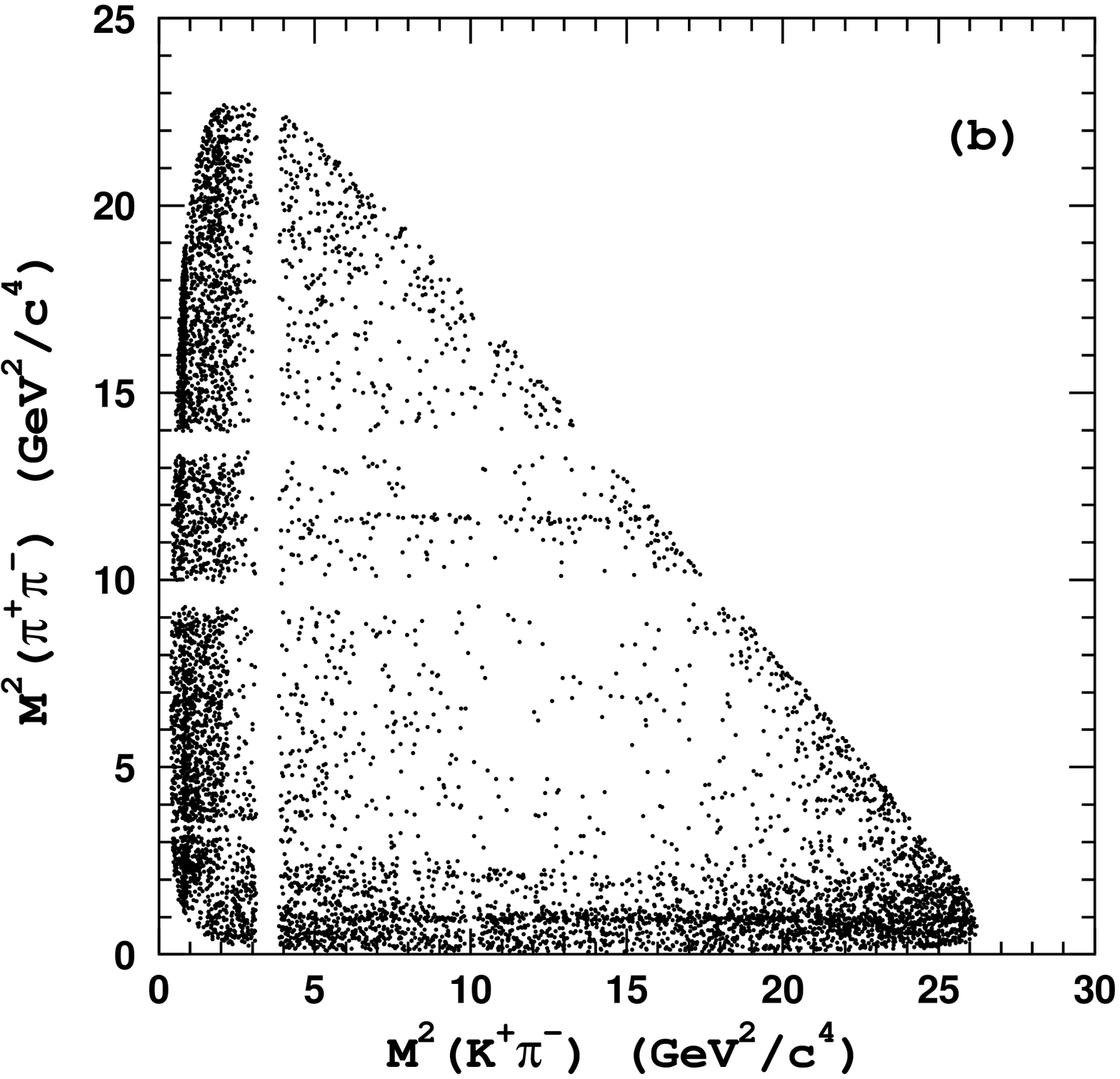}}  
\caption{\it a) The $B^+ \to K^+ \pi^+ \pi^-$ signal from Belle\cite{belle2}. b) The $B^+ \to K^+ \pi^+ \pi^-$
Dalitz plot.}
\label{bkpp}
\end{center}
\end{figure}

The $B^+ \to K^+ \pi^+ \pi^-$ signal and the Dalitz plot are shown in Fig. \ref{bkpp}. Fig. \ref{bkpp}-b 
illustrates how large is the phase space and how much the action is concentrated near the border.

\begin{figure}[H]
\begin{center}
 {\includegraphics[scale=0.29]{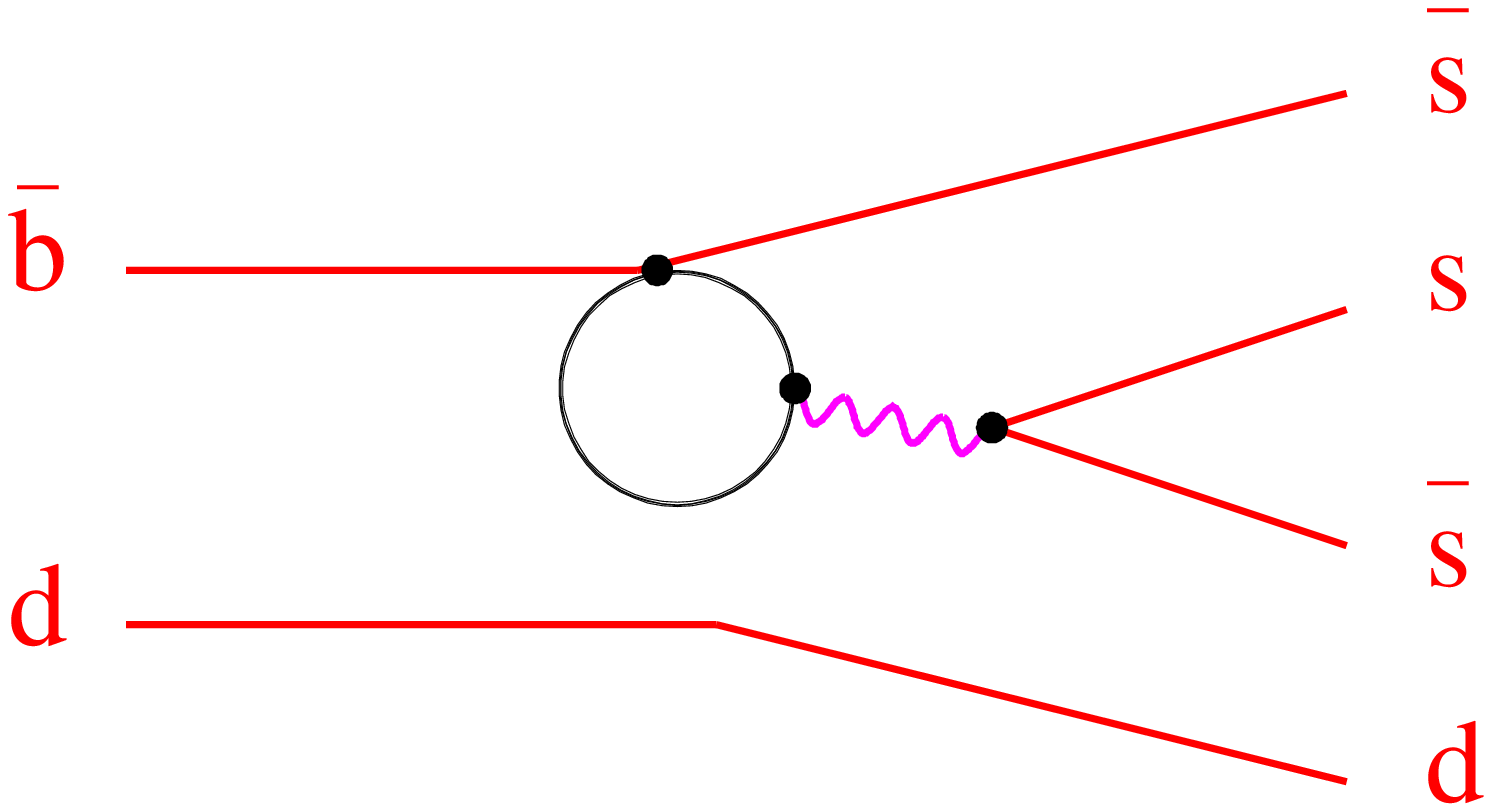}}
 {\includegraphics[scale=0.29]{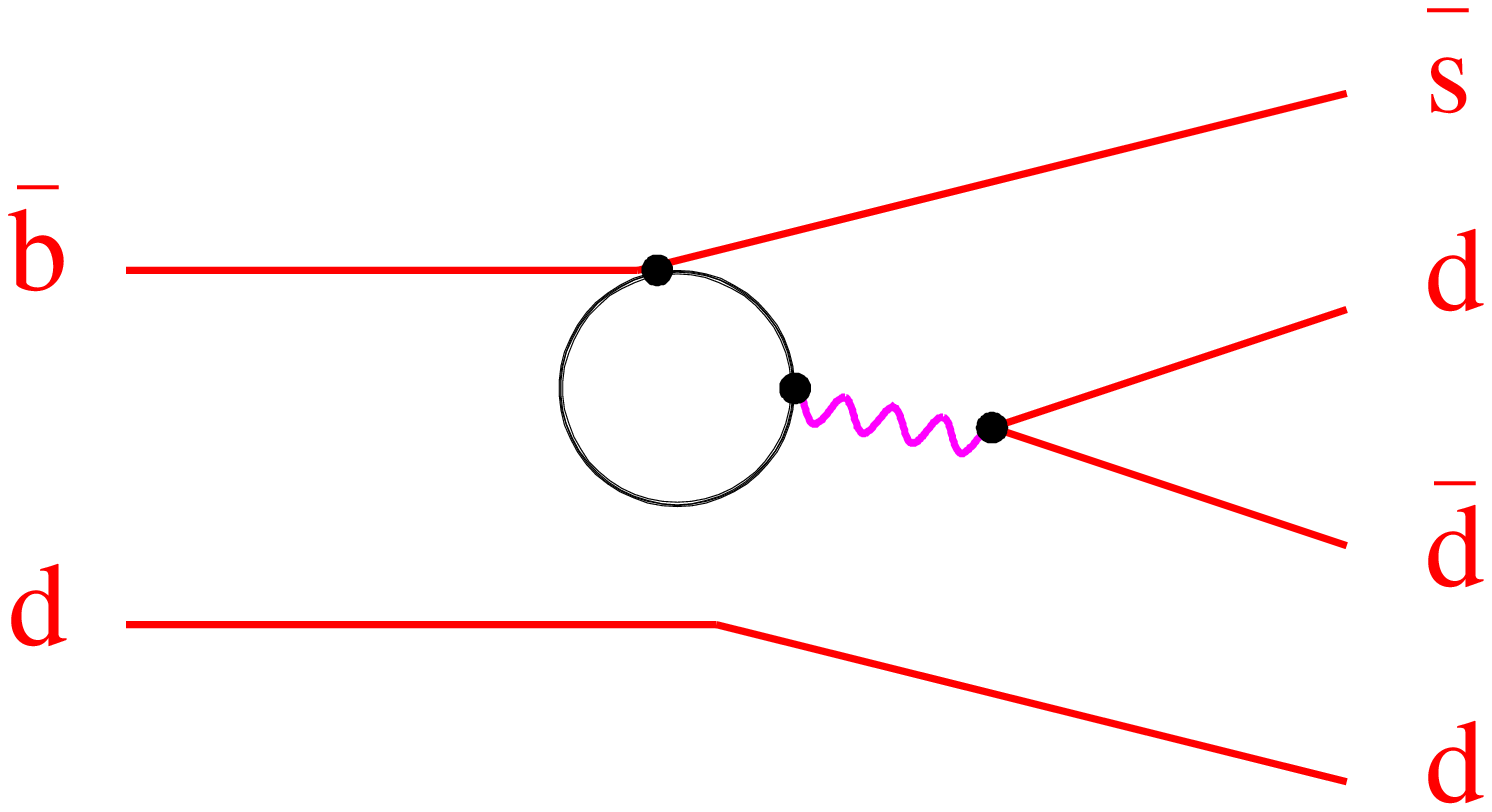}}  
\caption{\it a) Dominant diagram for $B \to KKK$ decays b) Dominant diagram for $B \to K \pi \pi$ decays.}
\label{penguin}
\end{center}
\end{figure}

In these decays the dominant mechanism for the $b\to s$ transition are penguin diagrams, shown in 
Fig. \ref{penguin}.  The diagram of Fig. \ref{penguin}-a leads to final states having three kaons. The
$K\pi\pi$ final states proceed via the diagram in Fig. \ref{penguin}-b. We expect dominant 
contributions from the $K^*(892) \pi^+$, $K^*_0(1430) \pi^+$, $\rho(770) K$ and $f_0K$ modes,
in addition to the nonresonant component. We also expect the decay fractions in both 
$B^0 \to K^0 \pi^+ \pi^-$ and $B^+\to K^+ \pi^+ \pi^-$ to be similar, since the replacement of the $d$ by 
the $u$ quark in Fig. \ref{penguin}-b turns the $B^0$ to the $B^+$ decay.

All studies of charmless hadronic three-body decays of $B$ mesons are performed with the isobar model.
The nonresonant is parameterized by empirical formulae, with independent coefficients for each axis. 
In the two $B \to K \pi \pi$ analyzes by Belle the expression for the nonresonant amplitude is
$\mathcal{A}_{NR} = a_1e^{i\delta_1}f(s_1) + a_2e^{i\delta_2}f(s_2)$. In both studies the data is better 
described by a model having one $\pi^+\pi^-$ scalar state at $m_{\pi\pi} \sim$ 1.5 GeV/c$^2$. The results 
of Dalitz plot fit are in Tables \ref{bplus} and \ref{b0}.

The decay fractions are similar in both decays, and correspond to the modes one expect from
the diagram of Fig. \ref{penguin}-b. In both $B^0$ and $B^+$ decays there is a large interference 
between the $K^*_0(1430) \pi^+$ and the $K\pi$ nonresonant component, which causes the fraction of the 
$K^*_0(1430) \pi^+$ to be very high. The $K\pi$ S-wave seems to be not well understood. 
The interference between the $\pi\pi$ S-wave and the corresponding nonresonant amplitude is small, though. 
The $f_0(X)$ state was represented by a Breit-Wigner function, whose parameters were determined by the data:
$m_0= 1.449 \pm 0.013$ GeV/c$^2$ and $\Gamma_0= 0.126 \pm 0.025$ GeV/c$^2$.
These values are in good agreement with the ones obtained from the $D^+_s \to \pi^-\pi^+\pi^+$ decay
by FOCUS and E791.

\begin{table}[t]
\centering
\caption{ \it Decay fractions (\%) of the $B^+ \to K^+\pi^+\pi^-$ decay, from a
fit using the isobar model for the S-wave amplitude.}
\begin{tabular}{|l|c|} \hline
 mode                              & fraction(\%)            \\
\hline
\hline
$K^*(892)^- \pi^+$      & 13.0$\pm$1.0   \\
$K^*_0(1430)^- \pi^+$   & 65.5$\pm$4.5   \\
$\rho(770)^0 K^+$       &  7.9$\pm$1.0   \\
$f_0(980) K^+$          & 17.7$\pm$3.6   \\
$f_0(X) K^+$            &  4.1$\pm$0.9   \\
$\mathrm{nonresonant}$  & 34.0$\pm$2.7   \\
\hline
\end{tabular}
\label{bplus}
\end{table}

\begin{table}[t]
\centering
\caption{ \it Decay fractions (\%) of the $B^0 \to K^0\pi^+\pi^-$ decay, from a
fit using the isobar model for the S-wave amplitude.}
\begin{tabular}{|l|c|} \hline
 mode                              & fraction(\%)            \\
\hline
\hline
$K^*(892)^- \pi^+$      & 11.8$\pm$1.7  \\
$K^*_0(1430)^- \pi^+$   & 64.8$\pm$7.8  \\
$\rho(770)^0 K^+$       & 12.9$\pm$2.0  \\
$f_0(980) K^+$          & 16.0$\pm$4.2  \\
$f_0(X) K^+$            &  3.7$\pm$2.4  \\
$\mathrm{nonresonant}$  & 41.9$\pm$5.5  \\
\hline
\end{tabular}
\label{b0}
\end{table}

It is interesting to these results to those from $B \to KKK$. From the diagram in Fig. \ref{penguin}-a
one expect significant contributions from the $\phi K$ and $f_0(980)K$. The nonresonant amplitude is 
parameterized by empirical formulae similar to those used in Belle analysis. In the case of the $B^0$  
the nonresonant amplitude has three independent terms. Like in the $B \to K\pi\pi$ decays, a scalar
$K^+K^-$ resonance was introduced, with mass and width determined by the fit.

\begin{figure}[H]
\begin{center}
 {\includegraphics[scale=0.29]{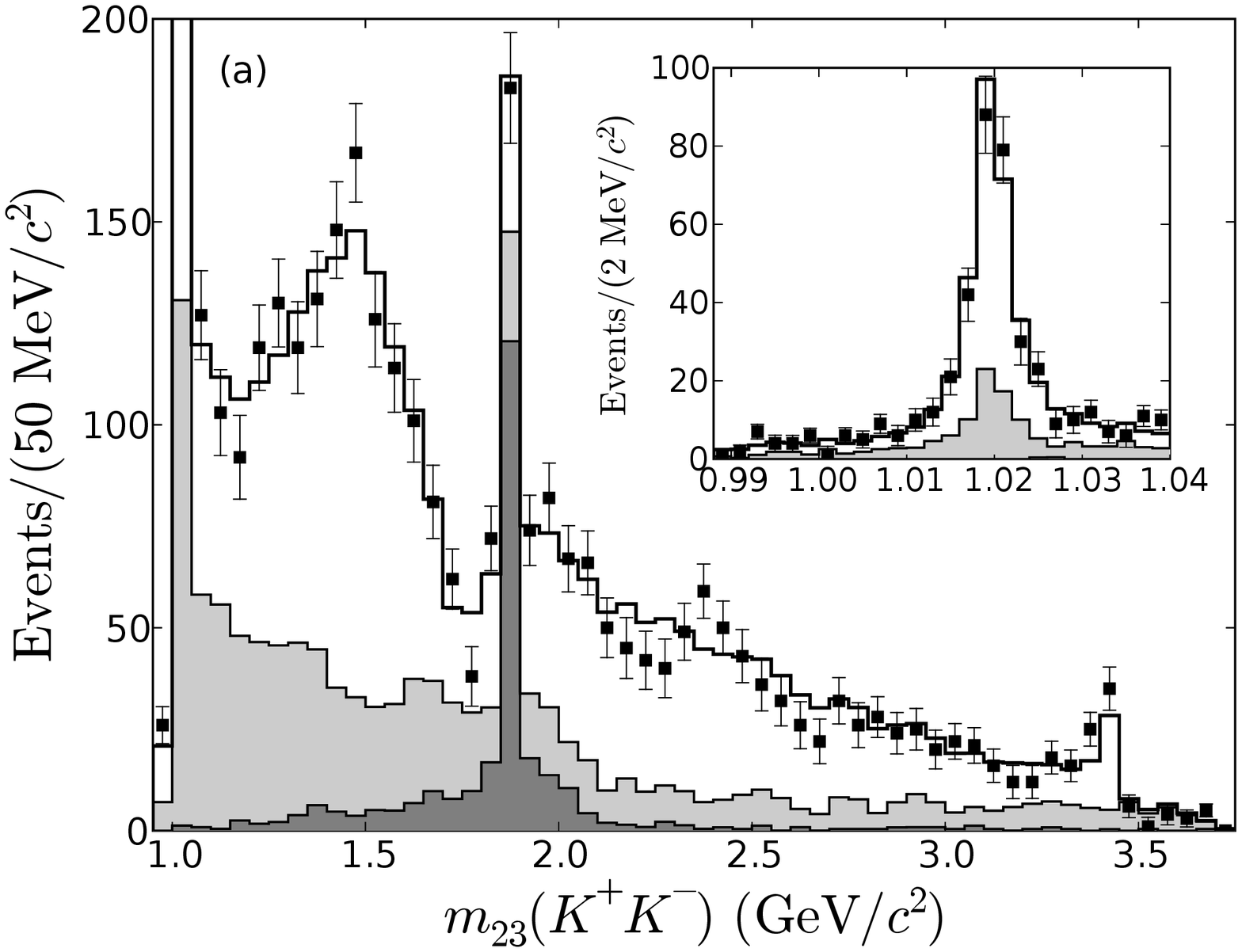}}
 {\includegraphics[scale=0.29]{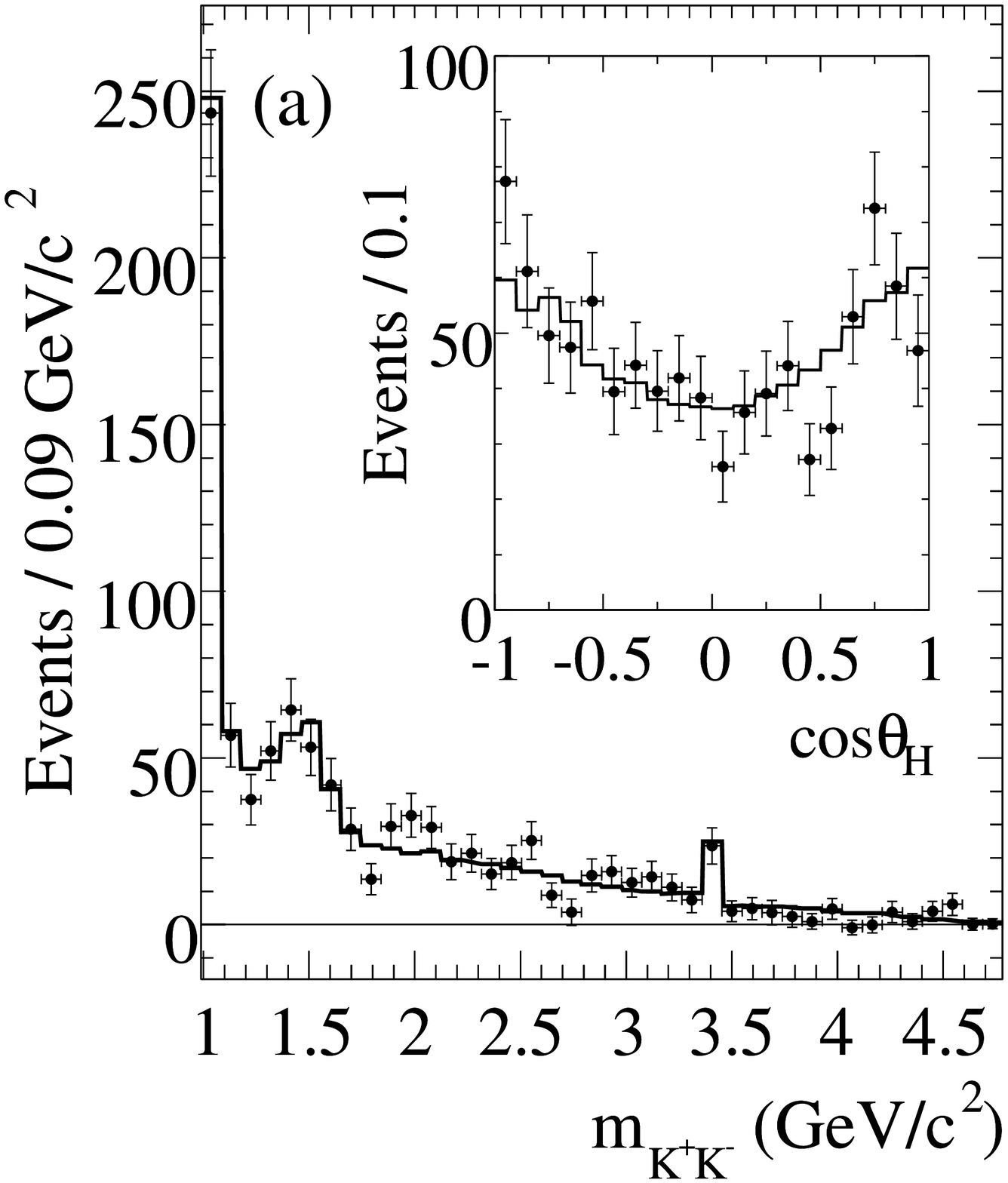}}  
\caption{\it a) The $K^+K^-$ projections from the $B^+ \to K^- K^+ K^-$ (left plot) and
 $B^0 \to K^0 K^+ K^-$ (right plot) Dalitz plots.}
\label{bkkk}
\end{center}
\end{figure}

The $K^+K^-$ projections of the Dalitz plot are shown in Fig. \ref{bkkk}. We see a clear bump next to the 
$\phi$ peak, at $m_{KK}\sim$1.5 GeV/c$^2$. The fit results are shown in Tables \ref{bpkkk} and \ref{b0kkk}.
The contribution of the $\phi K$ decay is similar in both cases. As in the case of the $D_s^+ \to K^+K^-\pi^+$
decay, the fraction of the $f_0(980) K$ mode suffers from large uncertainties. There is a small contribution
from the $f_0(1710) K$, but only in the $B^+$ decay. The most striking features, though, are the very different 
$K^+K^-$ S-wave composition and the enormous interference between the $K^+K^-$ nonresonant term and what BaBar 
calls the $X_0(1550)$ state. In the $B^+$ fit the sum of decay fractions amounts to 300\%!. 

The Breit-Wigner parameters of the $X_0(1550)$ state were determined in the $B^+$ analysis. In the $B^0$ analysis 
the $X_0(1550)$ parameters were fixed at the values obtained in the $B^+$ analysis. These are:
$m_0=1.539\pm0.020$ and $\Gamma_0 = 0.257 \pm 0.033$. 
GeV/c$^2$. The $X_0(1550)$ parameters are very different from the ones of the $f_0(X)$ state in $B \to K\pi\pi$.
However, before drawing any definitive conclusions, the $K^+K^-$ S-wave is understood, as well as the role and
form of the nonresonant amplitude.

\begin{table}[t]
\centering
\caption{ \it Decay fractions (\%) of the $B^+ \to K^- K^+ K^-$ decay, from a
fit using the isobar model for the S-wave amplitude.}
\begin{tabular}{|l|c|} \hline
 mode                              & fraction(\%)            \\
\hline
\hline
$\phi K^+$              & 11.8$\pm$1.2   \\
$f_0(980) K^+$          & 19$\pm$8       \\
$X_0(1550) K^+$         & 121$\pm$20     \\
$f_0(1710) K^+$         &  4.8$\pm$2.9   \\
$\mathrm{nonresonant}$  & 141$\pm$17     \\
\hline
\end{tabular}
\label{bpkkk}
\end{table}

\begin{table}[t]
\centering
\caption{ \it Decay fractions (\%) of the $B^0 \to K^0 K^+ K^-$ decay, from a
fit using the isobar model for the S-wave amplitude.}
\begin{tabular}{|l|c|} \hline
 mode                              & fraction(\%)            \\
\hline
\hline
$\phi K^0$              & 12.5$\pm$1.3  \\
$f_0(980) K^0$          & 40.2$\pm$9.6  \\
$X_0(1550) K^0$         & 4.1$\pm$1.3   \\
$f_0(1710) K^0$         &  -            \\
$\mathrm{nonresonant}$  & 112$\pm$15    \\
\hline
\end{tabular}
\label{b0kkk}
\end{table}

\section{Summary and conclusions}

Two of the most challenging problems in the scalar mesons physics have been discussed from the
point of view of heavy flavor decays. 

In the low energy $K\pi$ spectrum, the neutral $\kappa$ 
is now established. An analysis of the elastic scattering data revealed the position of the 
neutral $\kappa$ pole, in spite of the lack of data bellow 825 GeV/c$^2$, and of the suppression 
of the amplitude due to the Adler zero. The existence of the $\kappa$ charged partners remain 
unsettled, though. These issues can be addressed by heavy flavor decays. The $K\pi$ S-wave 
amplitude was measured in the  $D^+ \to K^-\pi^+\pi^+$ decay, including the region bellow 825 
GeV/c$^2$ where LASS data starts. There is, however, one unavoidable task: to extract the $K\pi$ 
elastic scattering phase from the measured amplitude one has to handle other strong 
interaction effects.

Evidences for a charged $\kappa$ are still scarce. The recent results from BaBar 
($D^0 \to K^-K^+\pi^0$) and Belle ($\tau^- \to \overline{K}^0\pi^- \nu_{\tau}$) 
are rather intriguing. In the $\tau$ decay the $K\pi$ system is isolated from any other strong 
interaction. We would expect that the $K\pi$ phase from this decay to match that of LASS, 
whereas in the case of the $D^0$ decay the three-body FSI, or a complex production amplitude,
would cause some deviations from the pure elastic scattering phase. The experimental results, 
however, show exactly the opposite picture. More data and refined analysis techniques are
clearly necessary. In the case of the$ \tau$ decay, the angular analysis is the crucial and 
missing piece. If such analysis is performed and confirm the resonant behavior at
low $K\pi$ mass, the we will have a compelling evidence for the charged $\kappa$.

In the $\pi\pi$ system, the existence of the $f_0(1370)$ has been also addressed in studies 
of HF decays. The analysis of the
$D^+_s \to \pi^-\pi^+\pi^+$ and $B \to K\pi\pi$ decays show that only one scalar state with mass
near 1.5 GeV/c$^2$ is necessary to describe the data. The measured parameters of this $f_0(X)$ 
state are consistent with those of the $f_0(1500)$. If we exclude the scalar mesons, in 
three body $D$ decays the only intermediate states observed are those having  a regular 
$q\bar q$ resonance. We could say that it is very likely that the $f_0(1500)$ is a genuine 
$q\bar q$ meson, or has, at least, a strong $q\bar q$ component in its wave function. On the
other hand, the evidence for the  $f_0(1370)$ in HF decays is weak and inconclusive. 
The puzzling fact, though, is that the fraction of the $f_0(1500)\pi^+$ mode in the 
$D^+_s \to \pi^-\pi^+\pi^+$ decay is very large, suggests that the $f_0(1500)$  has a
fairly large $s \bar s$ component and, therefore, a significant coupling to $KK$. But this is
not true for the $f_0(1500)$, and no indication of this state was found in 
$D^+_s \to K^-K^+\pi^+$ decay. One possible interpretation is that the S-wave model used in
the $D^+_s \to \pi^-\pi^+\pi^+$ analysis is incomplete.
There are also indications of a scalar state in $B \to KKK$ with mass near
1.5 GeV/c$^2$, but no conclusions can be drawn before the nonresonant component is understood.

We have seen that hadronic decays of heavy flavor are a very rich environment for the study of 
the scalar mesons. Thanks to their unique features, the information provided by 
HF decays are complementary  to the traditional hadronic collisions. One must also keep in mind 
that HF decays have been the only new data available in the past ten years, and this will be so
until the commissioning of the new facilities. The B-factories already have very large and clean 
samples of $D \to h_1h_2h_3$ decays. As for the $B \to h_1h_2h_3$, data with equivalent quality
will be available in a few years, from  LHCb and the other LHC experiments. 

The existence of good data, however, is not enough. The experimentalists need to develop better
analysis techniques, going beyond the isobar model. The limitations of the latter appear either 
when one moves to really high statistics, or when complex final states are analyzed. Even with a
better understanding of the S-wave, it seems that a simple coherent sum of amplitudes may be
too simplistic. The improvement of the analysis techniques does not depend only on the
experimentalists creativity, but also on a deeper understanding of the decay dynamics, of the role 
of final state interactions, of the nonresonant amplitude, form factors and line shapes. This
requires the urgent intervention of the theoreticians.

In high energy physics most of the attentions are turned to the searches for new physics.
There is a widespread belief that we are on the verge of great discoveries, that new particles
are right at the corner. One of the most promising fields is the phenomenon of CP violation. 
The correct measurement of the CP violation effects, however, depends on the accurate 
understanding of the low energy strong interaction dynamics. That's where the flavor physics 
and the hadron physics communities meet. Even with somewhat limited analysis tools, there are 
plenty of good  data from the B-factories that should be analyzed in a systematic way, with the 
focus on the physics of the light quarks and, in particular, of the scalar mesons. This is a 
very rich and challenging program that needs to be implemented.


\begin{thebibliography}{99}


\bibitem{asner} Particle Data Group, D. Asner {\it et al.}, J.Phys. G: Nucl.
Part. Phys. {\bf 33},716 (2006)

\bibitem{e791dp} E.M.Aitala {\it et al.} (E791 Collaboration), Phys. Rev. Lett. {\bf 86}, 770 (2001)

\bibitem{cg} E.M.Aitala {\it et al.} (E791 Collaboration), Phys. Rev. Lett. {\bf 89}, 121801 (2002)

\bibitem{meissner} S.Gardner and U.G.Meissner, Phys. Rev. D {\bf 65}, 094004 (2002).

\bibitem{sm} J.M.Link {\it et al.} (FOCUS Collaboration), Phys. Lett {\bf B585}, 200 (2004).

\bibitem{bm} E.M.Aitala {\it et al.} (E791 Collaboration), Phys. Rev.{\bf D73}, 032004 (2006).

\bibitem{beskappa} M.Abiklim {\it et al.} (BES Collaboration), hep-ex/0506055.

\bibitem{descortes} S.Descotes-Genon and B.Moussallam, Eur. Phys. J. C{\bf 48}, 553 (2006).

\bibitem{babar1} Babar Collaboration internal note, unpublished.

\bibitem{lass} D.Aston {\it et al.} (LASS Collaboration), Nucl. Phys. B{\bf 296}, 493 (1988).

\bibitem{kmat} J.M.Link {\it et al.} (FOCUS Collaboration), Phys. Lett {\bf B653}, 1  (2007).

\bibitem{belletau} D.Epifanov {\it et al.} (Belle Collaboration), arXiv:0706.2231.

\bibitem{massa} J.M.Link {\it et al.} (FOCUS Collaboration), Phys. Lett {\bf B621}, 72 (2005).

\bibitem{cleoc} G.Bonvicini {\it et al.} (CLEOc Collaboration), arXiv:0707.3060.

\bibitem{eu} FOCUS Collaboration, in preparation.

\bibitem{pdg} Particle Data Group, W.-M.Yao {\it et al.}, J.Phys. G: Nucl.
Part. Phys. {\bf 33},1 (2006).

\bibitem{bugg} D.V.Bugg, arXiv:0706.1254.

\bibitem{besf0} M.Abiklim {\it et al.} (BES Collaboration), Phys. Lett {\bf B607}, 243 (2005).

\bibitem{dsdp} FOCUS Collaboration internal note, unpublished.

\bibitem{e791ds} E.M.Aitala {\it et al.} (E791 Collaboration), Phys. Rev. Lett. {\bf 86}, 765 (2001).

\bibitem{babarkkp} B.Aubert  {\it et al.} (Babar Collaboration), Phys. Rev. {\bf D76}, 011102 (2007).

\bibitem{bellekspipi} L.M.Zhang {\it et al.} (Belle Collaboration), arXiv:0704.1000.

\bibitem{bed} I.Bediaga {\it et al.}, arXiv:0709.0075.

\bibitem{belle2} A. Garmash  {\it et al.} (Belle Collaboration), Phys. Rev. {\bf D71}, 092003 (2005).

\bibitem{belle3} A. Garmash  {\it et al.} (Belle Collaboration), Phys. Rev. {\bf D75}, 012006 (2007).

\bibitem{babarkkk1} B.Aubert  {\it et al.} (Babar Collaboration), Phys. Rev. {\bf D74}, 032003 (2006).

\bibitem{babarkkk2} B.Aubert  {\it et al.} (Babar Collaboration), arXiv:0706.3885.

\end{thebibliography}
\end{document}